\documentclass[a4,twocolumn]{IEEEtran}
\usepackage[ruled,vlined,norelsize]{algorithm2e}
\usepackage{tikz,color}
\usepackage{amsmath,amsfonts,amssymb,mathrsfs}
\usetikzlibrary{automata, positioning, matrix}
\usepackage{hyperref}
\usepackage{mathtools}
\usepackage{cite}
\usepackage{setspace}
\usepackage{hyperref}
\usetikzlibrary{arrows,automata}
\usetikzlibrary{decorations.pathmorphing}
\usetikzlibrary{shapes, arrows}
\usetikzlibrary{backgrounds}
\usetikzlibrary{fit}
\usetikzlibrary{petri}
\newtheorem{thm}{Theorem}
\newtheorem{mydef}{Definition}
\newtheorem{myeg}{Example}
\newtheorem{myrek}{Remark}
\newtheorem{mycom}{Complexity Analysis}
\newtheorem{myprb}{Problem}
\newtheorem{lem}{Lemma}
\usepackage{subcaption}

\title{Synthesis of  State-Attack Strategies for Anonymity and Opacity Violation in Discrete Event Systems}

\author{Xiaoyan Li and Christoforos N. Hadjicostis,~\IEEEmembership{Fellow,~IEEE} 
\thanks{This work was supported in part by  the Fundamental Research Program of Shanxi Province under Grant No. 202403021222156, in part by the Scientific and Technological Innovation Programs of Higher Education Institutions in Shanxi  under Grant No. 2024L178, in part by Shanxi Key Laboratory of Intelligent Detection Technology \& Equipment, North University of China, 030051, Shanxi Taiyuan, China, and in part by the China Scholarship Council.
(\it{Corresponding author: Christoforos N. Hadjicostis}.) }
\thanks{Xiaoyan Li is with the School of Information and Communication Engineering, North University of China, Taiyuan 030051, China (e-mail: 20220181@nuc.edu.cn).}
\thanks{Christoforos N. Hadjicostis is with the Department of Electrical and Computer Engineering, University of Cyprus, Nicosia, Cyprus (e-mail: hadjicostis.christoforos@ucy.ac.cy).}
 }

\begin{document}
\maketitle

\begin{abstract}
Attacks, including the manipulation of  sensor readings and the modification of  actuator commands,  pose a significant challenge to the security and privacy of automated systems.  
This paper considers discrete event systems that can be modeled with nondeterministic finite state automata that are susceptible to state attacks.
A state attack allows an intruder to learn whether or not the current state of a system falls into certain subsets of states.
The intruder has a limited total number of state attacks at its disposal, but can launch state attacks at arbitrary instants of its choosing. 
We are interested on violations of current-state anonymity (resp. opacity), i.e., situations where the intruder, based on the sequence of observations generated by the system and the outcome of any performed state attacks, can ascertain the exact current state of the system (resp. that the current state of the system definitely resides in a subset of secret states).
When the system violates current-state anonymity (resp. opacity) under a bounded number of state attacks,
a subsequent question is whether the intruder can design an attack strategy such that anonymity-violating (resp. opacity-violating) situations will always be reached.   
In this latter case, we also design an attack strategy that guarantees that the system will reach a violating   situation regardless of  system actions.
We provide pertinent complexity analysis of the corresponding verification algorithms and examples to illustrate the proposed methods. 
\end{abstract}


\section{Introduction}

Due to the integration of networking, computation, and physical processes, cyber-physical systems have been extensively adopted in various of applications, such as power systems, automated factories, autonomous vehicles, and flight control systems \cite{Sayed-2018,Alguliyev-2018,Duo-2022}. 
In general, the physical components of a system include sensors, actuators, and devices necessary  for the monitoring and control of underlying processes \cite{Sayed-2018,Alguliyev-2018}.
The controller receives sensor readings and processes data, based on which, it issues control commands to activate actuators and make the physical plant evolve as desired \cite{Sayed-2018,Alguliyev-2018}.
As the data between the controller and the sensors and actuators are transmitted via a network,
the integrity of the system may be comprised by  adversaries that manage to modify the transmitted data in a way that damages the system's normal operation (e.g., man-in-the-middle attacks) \cite{Sayed-2018,Alguliyev-2018,Duo-2022}.

Discrete event systems have been increasingly used to model cyber-physical systems from the lens of a logical abstraction \cite{Seatzu-2012,Mahulea-2020,Cassandras-2021,Hadjicostis-2020}.
In the last decades, in the realm of discrete event systems, many works have concentrated on cyber-attack-related problems \cite{Basilio-2021}, for instance, attack estimation \cite{He-2024}, state estimation under attack \cite{qi-2021,Yuting-2023},  attack detection and localization \cite{gao-2022,Fritz-2023}, and robust supervisor synthesis \cite{Wakaiki-2019,Alves-2022,Ma-2022,Meira-2021,Meira-2023,Wonham-2018}.

In order to keep a controlled system within the originally defined behavior under actuator enablement attacks, He {\it et al.} \cite{He-2024} devise  a reverse sensor module  to mislead an intruder's attack estimation.
Assuming that a malicious attacker is able to modify the sequence of observations generated by the system via insertion or erasure,
the work in \cite{qi-2021} addresses the joint state estimation under sensor attacks by constructing a joint estimator. In \cite{Yuting-2023}, current-state estimation under sensor attacks with cost constraints is addressed.
Regarding sensor attacks with one or more attack dictionaries,  an approach is proposed  in \cite{gao-2022} to determine whether a system has been attacked and which attack dictionaries have been used by the attacker, via a structure describing the observations generated by a system under attack. 
In \cite{Fritz-2023}, the models of replay attacks and covert attacks are presented, based of which detection and localization approaches for the corresponding attacks are proposed.

When there are multiple attackers that can perform sensor attacks to make the system generate a string outside a specific (given) language, Wakaiki {\it et al.}  in \cite{Wakaiki-2019} tackle the existence and the construction of a supervisor that aims to keep the system from generating behaviors that do not belong to the given language
when not knowing the exact attacker.
Following the same framework, the work in \cite{Alves-2022} presents  an approach for checking if a given language is P-observable for the existence of one attacker.  A  robust  supervisor is formulated and synthesized to handle a class of edit attacks on the sensor readings in \cite{Meira-2021}.  
Regarding actuator attacks that can re-enable events that are already disabled by the supervisor, 
the synthesis of a robust controller handling infinite actuator attacks  is investigated in  \cite{Ma-2022}, assuming that every unsafe state can be associated with a safety level represented by a positive integer. 
Furthermore, a robust supervisor  coping with both sensor and actuator attacks is synthesized in \cite{Meira-2023}. 

Unlike existing works addressing event attacks (including sensor and actuator attacks), this paper focuses on state attacks that can provide additional information about the system states to an intruder. 
Motivated by the limited intruder resources in practical applications, the number of state attacks that the intruder can perform is taken to be bounded.
The problem studied in this article relates to current-state estimation from the perspective of the intruder under a bounded number of state attacks. 
Note that an intelligent intruder may strategically decide {\em when} to  perform state attacks to achieve its objectives.   

Current-state estimation problems in the presence of attacks include the verification of current-state anonymity \cite{location_privacy} and current-state opacity \cite{Jacob-2016,Lafortune:2018,Li-2023}.
Current-state anonymity (resp. opacity) investigates whether the intruder can infer without deniability the exact current state of the system (resp. the membership of the possible current states to a given set of secret states).
Studying current-state estimation under a bounded number of state attacks, the main contributions in this article are listed below:
\begin{enumerate}
    \item The notion of current-state anonymity (resp. opacity) under a bounded number of state attacks is proposed to examine whether an intruder can determine the exact current state of the system (resp. the membership of the current state to a given subset of secret states) based on the sequence of observations generated by the system and any performed state attacks. 
In order to verify whether a given system satisfies current-state anonymity (resp. opacity) under a bounded number of state attacks, a verifier is constructed, based on which, an algorithm for the verification is developed.
    \item If a system is determined to violate current-state anonymity (resp. opacity) under a bounded number of state attacks, the notion of attack-enforced anonymity (resp. opacity) violation is presented to determine whether the intruder, under the given bounded number of state attacks, can {\em always} enforce a current-state anonymity (resp. opacity) violation.
A necessary and sufficient condition is derived for the verification of  attack-enforced anonymity (resp. opacity) violation; pertinent complexity analysis is also provided.
\item If a system is verified to be attack-enforced anonymity-violating  (resp. opacity-violating) under a bounded number of state attacks,  a feasible attack strategy is synthesized. The designed strategy to enforce current-state anonymity (resp. opacity) violation describes  attack decisions that  the intruder needs to make as observations are generated by the system.
\end{enumerate}

\section{Preliminaries}

Given an alphabet $E$, the set of all finite-length strings of elements in $E$ is denoted by $E^*$.  
Specifically, the empty string $\varepsilon$ belongs to $E^*$, i.e., $\varepsilon \in E^*$.
The concatenation of two strings $s_1, s_2\in E^{*}$, denoted by $s_1s_2$, describes the sequence of events captured by $s_1$ followed immediately by the sequence of events captured by $s_2$.
Given $s\in E^{*}$, if there exist $s',s''\in E^{*}$ such that $s=s's''$ holds,  $s'$ is said to be a prefix of  string $s$, denoted by $s'=s/s''$.  The set of all prefixes of $s$ is denoted by $\overline{s}$.
The length of a string $s$, denoted by $|s|$, is the number of all events in $s$. Note that for the empty string $\varepsilon$, we have $|\varepsilon|=0$.
Given a string $s$,
we use $s[i]\in E\cup \{\varepsilon\}$ to denote the $i$-th event, where $i\in \{0,1,2,\cdots,|s|\}$ (note that $s[0]=\varepsilon$).

A nondeterministic finite automaton (NFA) is defined as a quadruple ${G=(X, E, \delta, X_0)}$, where $X$ is the set of states, $E$ is the set of events, $\delta: X \times E \rightarrow 2^X$ is the transition function, and $X_0 \subseteq X$ is the set of initial states.
If (i) for all $x\in X$ and all $e\in E$, it holds that $|\delta(x,e)|\leq 1$  (in this scenario, $\delta$ becomes a deterministic, possibly partially defined, transition function, defined as $f: X\times E\rightarrow X$), and (ii) the initial state is unique (i.e., $|X_0|=1$), the NFA can be regarded as a deterministic finite automaton (DFA). 
To be clear, in the remainder of this paper, the transition function is denoted by $\delta$ if the considered automaton is an NFA, otherwise it is denoted by $f$ (i.e., if the considered automaton is a DFA).

The transition function $\delta$ can be extended from $X \times E\rightarrow 2^X$ to $X \times E^{*}\rightarrow 2^X$ in a recursive way, defined as $\delta(x, es)= \bigcup_{x'\in\delta(x, e)}\delta(x', s)$, where $e\in E$ and $s\in E^{*}$ (specifically, $\delta(x, \varepsilon)=\{x\}$ for all $x\in X$).
Given a set of states $X$ and an event $e$, the set of states that can be reached from $X$ via the occurrence of $e$ is denoted by $\delta(X, e)= \bigcup_{x\in X}\delta(x, e)$.
The behavior of a given NFA is its generated language, defined as $L(G)=\{s\in E^{*}|(\exists x_0\in X_0)\delta(x_0, s)\neq \emptyset\}$.
We also define the system behavior starting from a specific initial state $x_{0}\in X_{0}$ as $L(G, x_{0})=\{s\in E^{*}|\delta(x_0, s)\neq \emptyset\}$.
Given $x\in X$,  the set of events that are enabled at $x$ is denoted by $T(x)=\{e\in E\mid$ $\delta(x,e)\neq \emptyset\}$.
For a set of states $X_{1}\subseteq X$, $T(X_{1})=\{e\in E| (\exists x\in X_{1}) [\delta(x,e)\neq \emptyset]\}$ denotes the set of events that are enabled at $X_{1}$.
Given $s\in L(G)$, $s'$ is a suffix of $s$ if $ss'\in L(G)$; the set of all suffixes of $s$ is denoted by $SX(s)$.

In this paper, we consider (without loss of generality) fully-observable NFAs, i.e.,  all events involved are observable. 
Next, we give the formal definitions of current-state anonymity and opacity for fully-observable NFAs (in the absence of state attacks).

\begin{mydef}
  \label{def:current-state-anonymity} (Current-state anonymity \cite{location_privacy}) Given an NFA $G=(X,E,\delta,X_{0})$ with all events being observable, $G$ is said to be current-state anonymous if
  $$(\forall x_i\in X_{0}, \forall s\in L(G,x_i))[x\in \delta(x_i,s) \implies$$
  $$(\exists x_j\in X_{0},  \exists x'\neq x)[ x'\in \delta(x_j,s)]].$$
  $\hfill \square$
\end{mydef}

\begin{mydef}
  \label{def:current-state-opacity} (Current-state opacity \cite{Cassandras-2021}) Given an NFA $G=(X,E,\delta,X_{0})$ with all events being observable and a set of secret states $X_{S}\subset X$, $G$ is said to be current-state opaque with respect to $X_{S}$ if
  $$(\forall x_i\in X_{0}, \forall s\in L(G,x_i))~[ \delta(x_i,s)\cap X_{S}\neq \emptyset \implies$$
  $$(\exists x_j\in X_{0})~[ \delta(x_j,s)\cap X_{NS}\neq \emptyset]],$$ where $X_{NS}=X\setminus X_{S}$ is the set of non-secret states.
  $\hfill \square$
\end{mydef}

A well-known method for verifying current-state anonymity and opacity is to construct an observer.
We recall the formal definition of the observer and the verification condition below.

\begin{mydef}
    \label{def:observer}(Observer
    \cite{Hadjicostis-2020})
    Given a fully-observable NFA $G=(X,E,\delta,X_{0})$, the observer  of $G$ is a DFA $Obs(G)=(X_{obs}, E, f_{obs}, x_{0,{obs}})$, where
(i) $X_{obs}\subseteq 2^{X}$ is the set of states,
(ii)  $x_{0,obs}=X_{0}$ is the initial state, and
(iii) $f_{obs}: X_{obs}\times E\rightarrow X_{obs}$ is the transition function, defined as
$f_{obs}(x_{obs}, e)= \bigcup_{x\in x_{obs}}\delta(x, e)$ if 
 there exists $x\in x_{obs}$ such that $\delta(x, e)\neq \emptyset$ (otherwise, $f_{obs}(x_{obs}, e)$ is undefined). $\hfill \square$
\end{mydef}

\begin{thm}
    \label{thm:current-state-anonymity-verification}\cite{location_privacy}
    Consider an NFA $G=(X,E,\delta,X_{0})$ with all events being observable and its observer $Obs(G)=(X_{obs}, E, f_{obs}, x_{0,{obs}})$.
    The system $G$ is current-state anonymous if and only if 
    \[(\forall x_{obs}\in X_{obs})[|x_{obs}|>1].\]
 Given a set of secret states $X_{S}\subset X$, $G$ is current-state opaque if and only if
 \[(\forall x_{obs}\in X_{obs})[x_{obs}\cap X_{NS}\neq \emptyset],\]  where $X_{NS}=X\setminus X_{S}$ is the set of non-secret states.
    $\hfill \blacksquare$
   
\end{thm}

\begin{figure}[!htbp] 
\centering
\scalebox{0.7}{


\begin{tikzpicture}[->, node distance=1.5cm, on grid, auto]
  \node[state, initial] (1) {$1$};
  \node[state] (2) [right=of 1] {$2$};
  \node[state] (3) [below=of 2] {$3$};
  \node[state] (6) [above=of 2] {$6$};
  \node[state] (8) [right=of 6] {$8$};
  \node[state] (7) [above=of 8] {$7$};
  \node[state] (4) [right=of 2] {$4$};
  \node[state] (5) [below=of 4] {$5$};
  \node[state] (9) [above=of 6] {$9$};
  \node[state,initial] (10) [left=of 9] {$10$};

  \path[->]
    (1) edge node {$a$} (2)
        edge[bend right] node[left] {$a$} (3)
        edge[bend left=20] node[left] {$d$} (6)
         edge[bend left=20] node[left] {$d$} (9)
    (2) edge node {$b$} (4)
        edge node[right,yshift=-0.5cm,xshift=-0.5cm] {$c$} (5)
    (3) edge node[swap] {$b$} (5)
        edge node[right,yshift=0.5cm,xshift=-0.5cm] {$c$} (4)
    (4) edge[bend left] node {$c$} (5)
    (5) edge[swap] node {$c$} (4)
    (6) edge node {$b$} (7)
        edge node[swap] {$b$} (8)
    (7) edge[bend left] node {$c$} (8)
    (8) edge[swap] node {$c$} (7)
    (9) edge[swap] node[above] {$b$} (7)
    (10) edge[swap] node[above] {$d$} (9)
    (10) edge[swap] node[above] {$a$} (2);

      \node[state,rectangle, initial,right=of 7,xshift=2cm] (s1) {$\{1,10\}$};
    \node[state,rectangle] (s2) [below=of s1] {$\{2,3\}$};
    \node[state,rectangle] (s3) [below=of s2] {$\{4,5\}$};
    \node[state,rectangle] (s4) [right=of s2] {$\{6,9\}$};
    \node[state,rectangle] (s5) [below=of s4] {$\{7,8\}$};
    
    \path
    (s1) edge node {$a$} (s2)
    edge[bend left=20] node[above] {$d$} (s4)
    (s2) edge node {$b,c$} (s3)
    (s4) edge node {$b$} (s5)
    (s3) edge[loop below] node {$c$} ()
    (s5) edge[loop below] node {$c$} ();
\end{tikzpicture}

}
\caption{\label{fig:system}Fully-observable NFA and its observer.}
\end{figure}
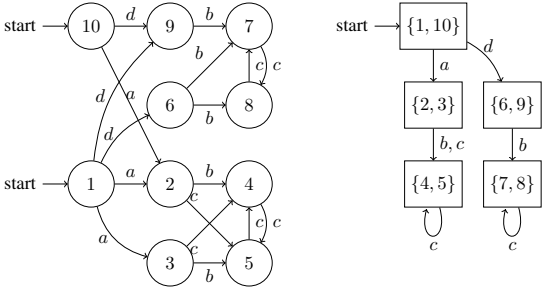

\begin{myeg}
    \label{eg:system-observer}
    Consider the NFA $G$ on the left of Fig.~\ref{fig:system}, where $E=\{a,b,c,d\}$ is the set of (observable) events, and $X_{0}=\{1,10\}$ is the set of initial states.
    Following Definition~\ref{def:observer}, the observer of $G$ is depicted on the right of Fig.~\ref{fig:system}. 
    By Theorem~\ref{thm:current-state-anonymity-verification}, $G$ is current-state anonymous since there is no singleton set in $Obs(G)$.
    Assuming that $X_{S}=\{7,8\}$, $G$ is not current-state opaque. If $X_{S}=\{2,5\}$, then $G$ is current-state opaque.
 $\hfill \triangle$\end{myeg}

We next review the composition operation of two automata since it is necessary to construct a composed structure to illustrate current-state estimation of the system at the intruder following the performed state attacks and obtained attack results as the system evolves. 

\begin{mydef}
    \label{def:compostion}  (Composition of two automata  \cite{Cassandras-2021})
     Consider two NFAs $G_{1}=(X_{1},E_{1},\delta_{1},X_{0,1})$ and $G_{2}=(X_{2},E_{2},\delta_{2}$, $X_{0,2})$. The composition of $G_{1}$ and $G_{2}$ is denoted by $G_{c}=G_{1} \bigoplus G_{2}=(X_{c},E_{c},\delta_{c},x_{0,c})$, where
    \begin{enumerate}
        \item $X_{c}=X_{1}\times X_{2}$ is the set of states;
        \item $E_{c}=E_{1}\cup E_{2}$ is the set of events;
        \item $x_{0,c}=X_{0,1}\times X_{0,2}$ is the set of initial states;
        \item $\delta_{c}$ is the transition function, defined as follows:
\begin{itemize}
    \item For $x_{c}=(x_{1},x_{2})$ and for $e\in E_{1}\cap E_{2}$, $\delta(x_{c},e)=\delta_{1}(x_{1},e)\times \delta_{2}(x_{2},e)$ if $\delta_{1}(x_{1},e)$ and $\delta_{2}(x_{2},e)$ are both defined; otherwise, $\delta(x_{c},e)$ is undefined.
    \item For $x_{c}=(x_{1},x_{2})$ and for $e\in E_{1}\setminus E_{2}$, $\delta(x_{c},e)=\delta_{1}(x_{1},e)\times \{x_{2}\}$ if $\delta_{1}(x_{1},e)$ is defined; otherwise, $\delta(x_{c},e)$ is undefined.
    \item For $x_{c}=(x_{1},x_{2})$ and for $e\in E_{2}\setminus E_{1}$, $\delta(x_{c},e)=\{x_{1}\}\times \delta_{2}(x_{2},e)$ if $\delta_{2}(x_{2},e)$ is defined; otherwise, $\delta(x_{c},e)$ is undefined.
\end{itemize}
    \end{enumerate}
    $\hfill \square$
\end{mydef}

\section{Anonymity and opacity  under a bounded number of state attacks}

Under the framework of automata, we formalize in this section
a system attack model on a set of attacked states for the underlying system.
Assuming that a maximum number of attacks is available to the intruder, we obtain a game structure that captures the interaction between the intruder and the system in the presence of state attacks.
An observation model, called attack-observer, under a bounded number of state attacks for the intruder, is constructed.
Based on this observation model, 
we derive the notions of current-state anonymity (resp. opacity) violation and attack-enforced anonymity (resp. opacity)  violation for the underlying system under a bounded number of state attacks.

\subsection{Problem setup}

A state attack performed by an intruder is associated with a subset of states $X_{A}\subset X$. 
The intruder can decide when to perform a state attack, and 
if it launches a state attack, it can obtain an accurate (binary) result on whether or not the current state of the system is in $X_{A}$. 
Note that if the current state of the system is not in $X_{A}$, then it is in $X_{NA}=X\setminus X_{A}$.
We use event ``$Y$"  (resp. ``$N$") to represent that the intruder performs (resp. does not perform) a state attack, and events ``0" or ``1" to denote the result of the performed state attack, where ``0" denotes that the current state of the system is in $X_{NA}$, and event ``1" indicates that the current state of the system is in $X_{A}$. 
Motivated by intruder limitations in terms of  existing resources in the real-world applications, we assume that the number of state attacks that an intruder can perform is bounded.
We use $D\in \mathbb{N}$ to denote the resources that the intruder initially has, i.e., the maximum number of state attacks that the intruder can perform. 

In the absence of  state attacks, the intruder deduces the current state of the system based on the sequence of observations.
Intuitively, when the intruder exerts state attacks, it can fuse the sequence of observations generated by the system with the received attack results to more accurately infer the current state of the system.
Note that the intruder can decide whether or not to exert a state attack when a new observation is generated and is assumed to have a full knowledge of the system.
 However, the result of a state attack depends on the exact current state of the system, which introduces uncertainty when estimating the current state of the system.  
From perspective of the intruder,
based on the sequence of observations associated with performed state attacks and received attack results,
the following two problems arise.

\begin{myprb}
    Can the intruder, based on the sequence of observations generated by the system and any attack results of a bounded number of  performed state attacks, infer with certainty that the possible current-state estimate is a singleton set (for anonymity violation) or that it is a subset of the set of secret states (for current-state opacity violation)?
\end{myprb}

If the answer to Problem~1 is yes, the underlying system violates current-state anonymity (resp. opacity) under a bounded number of state attacks. Then, a subsequent problem becomes apparent as follows.

\begin{myprb}
    Can the intruder strategize (based on the information it has, namely the set of possible system states and the number of state attacks) so that it can always force the system to reach an anonymity-violating (resp. opacity-violating)  situation under the given bounded number of state attacks? 
\end{myprb}

If the answer to Problem~2 is yes, the underlying system is attack-enforced anonymity-violating (resp. attack-enforced opacity-violating) under a bounded number of state attacks.
Then, the intruder might be interested in designing a feasible attack policy to guarantee that an anonymity-violating (resp. opacity-violating) situation can be reached.

Based on the problem setup aforementioned, the focus of this paper is to check, given an underlying nondeterministic system, current-state anonymity (resp. opacity) violation,  attack-enforced anonymity (resp. opacity) violation, and the synthesis of a feasible attack strategy for attack-enforced anonymity (resp. opacity) violation.

\subsection{Problem formulation}
\begin{mydef}
    \label{def:system-attack}(System attack model)
    Consider an NFA $G=(X,E,\delta,X_{0})$ with all events being observable.  The system attack model of $G$ with respect to set of attacked states $X_{A}$, $X_{A}\subset X$, is denoted by an NFA $G_{a}=(X_{a}, E_{a},\delta_{a},X_{0,a})$, where
    \begin{enumerate}
        \item $X_{a}=X$ is the set of states;
        \item $E_{a}=E\cup E_{r}$  is the set of events, where $E_{r}=\{0,1 \}$;
        \item $\delta_{a}=\delta\cup \delta_{r}$ is the transition function, defined as:\\ (i) for all $x\in X_{A}$, $\delta_{r}(x,1)=\{x\}$, and\\ (ii) for all $x\in X\setminus X_{A}$,  $\delta_{r}(x,0)=\{x\}$;
        \item $X_{0,a}=X_{0}$ is the set of initial states.
    \end{enumerate}
     $\hfill \square$
\end{mydef}

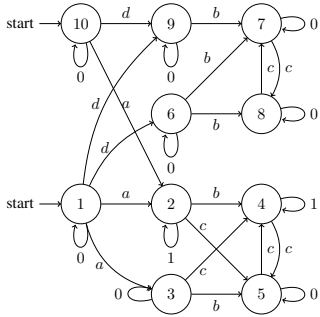
\begin{figure}[!htbp] 
\centering
\scalebox{0.6}{


\begin{tikzpicture}[->, node distance=2cm, on grid, auto]
  \node[state, initial] (1) {$1$};
  \node[state] (2) [right=of 1] {$2$};
  \node[state] (3) [below=of 2] {$3$};
  \node[state] (6) [above=of 2] {$6$};
  \node[state] (8) [right=of 6] {$8$};
  \node[state] (7) [above=of 8] {$7$};
  \node[state] (4) [right=of 2] {$4$};
  \node[state] (5) [below=of 4] {$5$};
   \node[state] (9) [above=of 6] {$9$};
   \node[state,initial] (10) [left=of 9] {$10$};

  \path[->]
    (1) edge node {$a$} (2)
        edge[bend right] node[left] {$a$} (3)
        edge[bend left=20] node[left] {$d$} (6)
    (2) edge node {$b$} (4)
        edge node[right,yshift=-0.5cm,xshift=-0.5cm] {$c$} (5)
    (3) edge node[swap] {$b$} (5)
        edge node[right,yshift=0.5cm,xshift=-0.5cm] {$c$} (4)
    (4) edge[bend left] node {$c$} (5)
    (5) edge[swap] node {$c$} (4)
    (6) edge node {$b$} (7)
        edge node[swap] {$b$} (8)
    (7) edge[bend left] node {$c$} (8)
    (8) edge[swap] node {$c$} (7)
  (1)  edge[bend left=20] node[left] {$d$} (9)
         (9) edge[swap] node[above] {$b$} (7);

  \path[->]
    (2) edge[loop below] node {$1$} (2)
    (4) edge[loop right] node {$1$} (4)
    (7) edge[loop right] node {$0$} (7)
    (1) edge[loop below] node {$0$} (1)
    (3) edge[loop left] node {$0$} (3)
    (5) edge[loop right] node {$0$} (5)
    (6) edge[loop below] node {$0$} (6)
     (9) edge[loop below] node {$0$} (9)
    (8) edge[loop right] node {$0$} (8)
    (10) edge[swap] node[above] {$d$} (9)
    (10) edge[swap] node[above] {$a$} (2)
      (10) edge[loop below] node {$0$} (10);

\end{tikzpicture}

}
\caption{\label{fig:system-attack} System attack model for the NFA in Fig.~\ref{fig:system} when $X_{A}=\{2,4\}$.}
\end{figure}

\begin{figure}[!htbp] 
\centering
\scalebox{0.6}{


\begin{tikzpicture}[->, node distance=2.2cm and 2.5cm, on grid, auto,
    state/.style={rectangle, draw, minimum height=0.8cm, minimum width=1cm}]
    
  \node[state, initial] (s1) {$\{1,10\}$};
  \node[state] (s23) [right=of s1] {$\{2,3\}$};
  \node[state] (s6) [above=of s23] {$\{6,9\}$};
  \node[state] (s45) [right=of s23] {$\{4,5\}$};
  \node[state] (s78) [right=of s6] {$\{7,8\}$};
  \node[state] (s2) [below=of s45] {$\{2\}$};
  \node[state] (s3) [below=of s23] {$\{3\}$};
  \node[state] (s4) [right=of s45] {$\{4\}$};
  \node[state] (s5) [below=of s4] {$\{5\}$};

  \path
    (s1) edge[loop below] node {$0$} ()
         edge node {$a$} (s23)
         edge[bend left] node[left] {$d$} (s6)
    (s23) edge[swap] node[above] {$b,c$} (s45)
          edge node[left] {$1$} (s2)
          edge node[right] {$0$} (s3)
    (s6) edge node[above] {$b$} (s78)
         edge[loop below] node {$0$} ()
    (s45) edge[loop above] node {$c$} ()
          edge node[above] {$1$} (s4)
          edge[bend left=15] node[above] {$0$} (s5)
    (s2) edge node[above] {$c$} (s5)
         edge[loop below] node {$1$} ()
         edge[bend right=15] node[left] {$b$} (s4)
    (s3) edge node[right] {$c$} (s4)
         edge[bend right=55] node[below] {$b$} (s5)
         edge[loop below] node {$0$} ()
    (s78) edge[loop right] node {$0,c$} ()
         
    (s4) edge[bend left=15] node[right] {$c$} (s5)
         edge[loop above] node {$1$} ()
    (s5) edge[swap] node[left] {$c$} (s4)
         edge[loop below] node {$0$} ();

\end{tikzpicture}

}
\caption{\label{fig:observer_system-attack} Observer for the NFA in Fig.~\ref{fig:system-attack}.}
\end{figure}
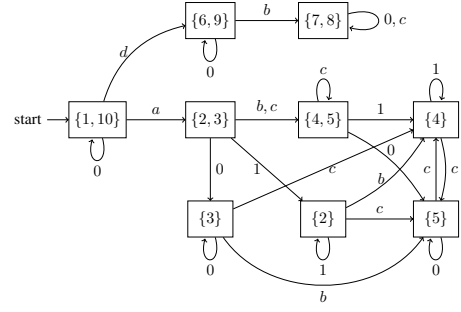

\begin{myeg}
    \label{eg:observer} (Example~\ref{eg:system-observer} continued)
    Reconsider the NFA $G$ in Fig.~\ref{fig:system}.
Assume that  $X_{A}=\{2,4\}$.   According to Definition~\ref{def:system-attack}, the system attack model $G_{a}$ of $G$ is obtained, as shown in Fig.~\ref{fig:system-attack}.
    Following Definition~\ref{def:observer}, the observer of  $G_{a}$ is visualized in Fig.~\ref{fig:observer_system-attack}.
 $\hfill \triangle$\end{myeg}


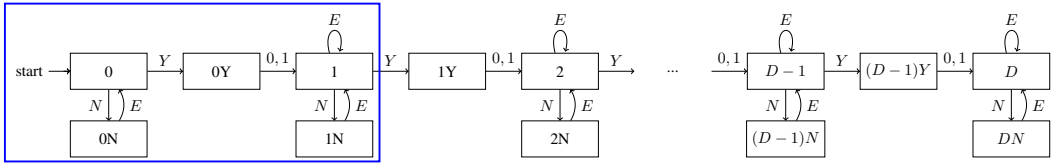
\begin{figure*}[!htbp] 
\centering
\scalebox{0.6}{
\begin{tikzpicture}[->, node distance=2.5cm and 2.5cm, on grid, auto,
    state/.style={rectangle, draw, minimum height=0.8cm, minimum width=1.7cm}]

			\node[state,initial] (0)   {0};
            \node[state,right of=0] (0Y)   {0Y};
            \node[state,below of=0,yshift=1cm] (0N)   {0N};
			\node[state] (1) [right of=0Y] {1};
            \node[state] (1Y) [right of=1] {1Y};
            \node[state] (1N)   [below of=1,yshift=1cm]{1N};
			\node[state] (2) [right of=1Y] {2};
            \node[state] (2N) [below of=2,yshift=1cm] {2N};
			\node[state,draw=white] (3) [right of=2] {...};
			\node[state] (4) [right of=3]  {$D-1$};
            \node[state] (4Y) [right of=4]  {$(D-1)Y$};
            \node[state] (4N) [below of=4,yshift=1cm]  {$(D-1)N$};
            \node[state] (5) [right of=4Y]  {$D$};
            \node[state] (5N) [below of=5,yshift=1cm]  {$DN$};

			\draw  (-1,0) -- (0);
            \path
            (0) edge   node[above] {$Y$}  (0Y)
             (0) edge   node[left] {$N$}  (0N)
             (0N) edge[bend right]   node[right] {$E$}  (0)
             (0Y) edge   node[above] {$0,1$}  (1)

              (1) edge   node[above] {$Y$}  (1Y)
             (1) edge   node[left] {$N$}  (1N)
             (1N) edge[bend right]   node[right] {$E$}  (1)
             (1Y) edge   node[above] {$0,1$}  (2)

              (2) edge   node[above] {$Y$}  (3)
             (2) edge   node[left] {$N$}  (2N)
             (2N) edge[bend right]   node[right] {$E$}  (2)
             (3) edge   node[above] {$0,1$}  (4)

              (4) edge   node[above] {$Y$}  (4Y)
             (4) edge   node[left] {$N$}  (4N)
             (4N) edge[bend right]   node[right] {$E$}  (4)
             (4Y) edge   node[above] {$0,1$}  (5)

              (5) edge   node[left] {$N$}  (5N)
              (5N) edge [bend right]node[right] {$E$} (5)
              (2) edge[loop above] node[above] {$E$} (2)
              (1) edge[loop above] node[above] {$E$} (1)
              (4) edge[loop above] node[above] {$E$} (4)
                (5) edge[loop above] node[above] {$E$} (5)
;
\draw[blue,very thick] (-2.3,1.5) rectangle (6,-2);
			
		\end{tikzpicture}
}
\caption{\label{fig:number-attack} Number attack model for maximum number of $D\in \mathbb{N}$ state attacks.}
\end{figure*}

\begin{mydef}
    \label{def:number-attack}(Number attack model)
     Consider an NFA $G=(X,E,\delta,X_{0})$ with all events being observable. 
     The number attack model for maximum $D$ attacks, is denoted by a DFA $G_{D}=(X_{D},E_{D},f_{D},x_{0,D})$, where 
    \begin{enumerate}
        \item $X_{D}=\{0,1,2,\cdots,D\}\cup \{0N,1N,2N,\cdots,DN\}\cup \{0Y,1Y,2Y,\cdots,(D-1)Y\}$ is the set of states;
        \item  $E_{D}=E\cup E_{A}\cup E_{r}$ is the set of events, where $E_{A}=\{Y,N\}$ and $E_{r}=\{0,1\}$;
        \item $f_{D}$ is the transition function, defined as:
\begin{itemize}
    \item for all $x_{i}\in \{0,1,2,\cdots,D-1\}$ and for $Y\in E_{A}$, $f_{D}(x_{i},Y)=x_{i}Y$,
    \item for all $x_{i}\in \{0,1,2,\cdots,D\}$ and for  $N\in  E_{A}$, $f_{D}(x_{i}, N)=x_{i}N$,
    \item for all $x_{i}\in \{1,2,\cdots,D\}$ and for all $e\in E\subseteq E_{D}$, $f_{D}(x_{i},e)=x_{i}$, 
    \item for all $x_{i}N\in \{0N,1N,2N,\cdots,DN\}$ and for all $e\in E\subseteq E_{D}$, $f_{D}(x_{i}N$, $e)=x_{i}$, and
    \item for all $x_{i}Y\in  \{0Y,1Y,2Y,\cdots,(D-1)Y\}$ and for all $\sigma\in E_{r}\subseteq E_{D}$, $f_{D}(x_{i}Y,\sigma)=x_{i+1}$;
\end{itemize}
\item $x_{0,D}=0$ is the initial state.
    \end{enumerate}$\hfill \square$
    
\end{mydef}

The number attack model in Definition~\ref{def:number-attack} is graphically represented in Fig.~\ref{fig:number-attack} (ignore the blue boundary for now). There are in total $3D+2$ states, and each state is labeled with $Y$, or $N$, or not labeled at all. 
Given $x\in \{0,1,\cdots, D\}$,  state $x$ (not labeled with $Y$ or $N$) represents the number of state attacks that have been executed by the intruder thus far;
the state $xN$ (labeled with $N$) represents that (i) the number of state attacks that have been executed by the intruder thus far is $x$ and (ii) the intruder is waiting for the system to generate  subsequent activity.
Given $x\in \{0,1,\cdots, D-1\}$, state $xY$ (labeled with $Y$) represents that the intruder is in the process of performing the $(x+1)$-st state attack. 

The occurrence of an event $N\in E_{A}$  at each state $x\in \{0,1,\cdots, D\}$  (not labeled with $Y$ or $N$) illustrates that the intruder does not perform a state attack; this will not increase the number of state attacks that are available to the intruder, denoted by $f_{D}(x, N)=xN$.
The occurrence of an event $e\in E$ at each state $x\in \{1,2,\cdots, D\}$ (not labeled with $N$ or $Y$)  illustrates that the system can execute any activity after the last state attack action is made; this will not increase the number of state attacks that are available to the intruder, denoted by $f_{D}(x, e)=x$ for all $e\in E$.
The occurrence of an event $e\in E$ at each state $x\in \{0N,1N,\cdots, DN\}$ (labeled with $N$)  illustrates that the system can execute any activity when no new state attack is performed; this will not increase the number of state attacks that are available to the intruder, denoted by $f_{D}(xN, e)=x$ for all $e\in E$.

The occurrence of an event $Y\in E_{A}$ at each state $x\in \{0,1,2,\cdots,D-1\}$ (not labeled with $Y$ or $N$) indicates that the intruder launches the $(x+1)$-st state attack; this will not increase the number of state attacks that have been performed by the intruder thus far, denoted by $f_{D}(x$, $Y)=xY$.
The occurrence of events $0$ and $1$ at each state $xY\in \{0Y,1Y,2Y,\cdots,(D-1)Y\}$ (labeled with $Y$) suggests that the intruder has obtained the result of the $(x+1)
$-st  state attack; this  implies that the $(x+1)
$-st state attack is completed and the number of attacks that are available to the intruder increases by $1$, denoted by  $f_{D}(x_{i}Y,\sigma)=x_{i+1}$, where $\sigma\in E_{r}$.

\begin{mydef}
    \label{def:game-structure-automaton}
 (Game-structure automaton)   
 Let $Y$ and $N$ be the attack-yes and attack-no events, respectively, and ``1" and ``0" be the attack result of a state attack.
 The game structure between the intruder and the system is denoted by
 $G_{si}=(X_{si},E_{si},f_{si},x_{0,si})$, where
    \begin{enumerate}
        \item $X_{si}=\{A,AY,S\}$ is the set of states,
        \item $E_{si}=E\cup \{Y,N,0,1\}$ is the set of events,
        \item $x_{0,si}=A$ is the initial state, and
        \item  $f_{si}$ is the transition function, defined as:
\begin{itemize}
    \item for $A$ and for $N$, $f_{si}(A,N)=S$;
    \item for $A$ and for $Y$, $f_{si}(A,Y)=AY$;
    \item for $AY$ and for all $\sigma\in \{0,1\}$ $f_{si}(AY,\sigma)=S$;
    \item for $S$ and for all $e\in E$, $f_{si}(S,e)=A$.
\end{itemize}
    \end{enumerate}
    $\hfill \square$
\end{mydef}

\begin{figure}[!htbp] 
\centering
\scalebox{0.7}{
\begin{tikzpicture}[->]
			
\node[state,initial,fill=green!10] (0)   {$A$};
\node[state,right of=0,xshift=1cm,fill=green!10] (1)   {$AY$};
    \node[state,right of=1,xshift=2cm,fill=red!10] (0Y)   {$S$};


    \path
        (0) edge[bend right]   node[above] {$N$}  (0Y)
        (1) edge   node[above] {$0,1$}  (0Y)
        (0Y) edge [bend right]  node[above] {$E$}  (0)
        (0) edge   node[above] {$Y$}  (1);

\end{tikzpicture}
}
\caption{\label{fig:system-intruder-game} Game structure between the intruder and the system in the presence of state attacks.}
\end{figure}
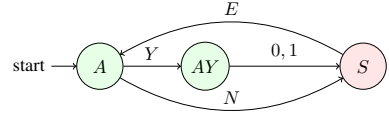

Game theory \cite{game-theory} is an appropriate  methodology framework to describe the interaction between the system and the intruder since there are two players in the problem setting.
The game structure between the system and intruder is graphically represented in Fig~\ref{fig:system-intruder-game}. 
There are three states: $A$ indicates that the current player is the intruder and is deciding whether to perform a state attack or not, $AY$ indicates that the current player is the intruder and is waiting for the attack result, and $S$ indicates that the current player is the system, which can also be regarded as the intruder inferring the situation of the system after the attack action.
At the initial state $A$, if the intruder does not perform a state attack, i.e., $N$ occurs and $S$ is reached, then it is the intruder's turn to infer the current state of the system and also the system's turn to generate subsequent activity.
If the intruder performs a state attack, i.e., $Y$ occurs, then it is still the intruder's turn to wait for the attack result ``0" or ``1", i.e., $AY$ is reached.
When the attack result is obtained,  i.e., ``0" or ``1" occurs and $S$ is reached, it is the system's turn to generate subsequent activity and the intruder's turn to obtain the state estimate of the system.
If the system generates a new event $e\in E$ at $S$, i.e, $A$ is reached, then it is the intruder's turn to again decide whether to perform a new state attack or not.

\begin{mydef}
    \label{def:system-intruder-game-attack}
     (Game-structure automaton under a bounded number of state attacks)   
     Consider an NFA $G=(X,E,\delta,X_{0})$ with all events being observable, under a bounded number of state attacks captured by $(X_{A},D)$, where $X_{A}$ is the set of attacked states and $D \in \mathbb{N}$ is the maximum number of attacks.
   Let  $G_{si}=(X_{si},E_{si},f_{si},x_{0,si})$ be the game structure between the intruder and the system under $E_{A}=\{Y,N\}$ and $E_{r}=\{0,1\}$ and  $G_{D}=(X_{D},E_{D},f_{D},x_{0,D})$ be the number attack model.
The game-structure automaton under a bounded number of state attacks is the composition of $G_{si}$ and $G_{D}$, denoted by $G_{siD}=(X_{siD},E_{siD},f_{siD},x_{0,siD})$. $\hfill \square$
\end{mydef}

{The game-structure automaton under a bounded number of $D$ state attacks is graphically represented in Fig.~\ref{fig:system-intruder-game-attack-D} (ignore the blue boundary for now).
Given a state $x_{siD}=(x_{siD}^{1},x_{siD}^{2})\in X_{siD}$, $x_{siD}^{1}$ denotes the situation of the intruder, and $x_{siD}^{2}$ denotes the situation of the state attacks that the intruder has performed. 
We use $x_{siD}^{2}(D)$ to denote the number of state attacks derived from $x_{siD}^{2}$. 
We consider the following scenarios:
(i) If $x_{siD}^{1}=A$ and $x_{siD}^{2}\in \{0,1,\cdots,D-1\}$, the intruder is deciding whether or not to perform the $(x_{siD}^{2}+1)$-st state attack; 
(ii) If $x_{siD}^{1}=AY$ and $x_{siD}^{2}\in \{0Y,1Y,\cdots,(D-1)Y\}$, the intruder has performed the $(x_{siD}^{2}(D)+1)$-st state attack and is waiting for the attack result; 
(iii) If $x_{siD}^{1}=S$ and $x_{siD}^{2}\in \{1,2,\cdots,D\}$, the intruder has obtained the attack result of the $x_{siD}^{2}$-th state attack;
(iv) If $x_{siD}^{1}=S$ and $x_{siD}^{2}\in \{0N,1N,\cdots,(D-1)N\}$, the intruder has decided not to perform the $(x_{siD}^{2}(D)+1)$-st state attack; 
(v) If $x_{siD}^{1}=A$ and $x_{siD}^{2}= D$, the intruder cannot perform a state attack anymore;
(vi) If $x_{siD}^{1}=S$ and $x_{siD}^{2}= DN$, the intruder does not perform a state attack (it cannot perform a state attack anymore).
Obviously, states with the characteristic ``$S$" are states at which the intruder attempts to update its state estimates of the system; these states are filled in red color.
\begin{figure}[!htbp] 
\centering
\scalebox{0.6}{
\begin{tikzpicture}[->, node distance=1.5cm, on grid, auto,
    state/.style={rectangle, draw, minimum height=0.5cm, minimum width=0.8cm}]
			
\node[state,initial,xshift=1cm,fill=green!10] (0)   {$A,0$};
    \node[state,right of=0,xshift=1cm,fill=red!10] (0Y)   {$S,0N$};
    
    \node[state,below of=0,fill=green!10] (2)   {$AY,0Y$};
    \node[state,below of=2,fill=green!10] (01)   {$A,1$};
    \node[state,fill=red!10,left of=01,xshift=-1cm] (s1)   {$S,1$};	
    \node[state,right of=01,xshift=1cm,fill=red!10] (0Y1)   {$S,1N$};
    
     \node[state,below of=01,fill=green!10] (3)   {$AY,1Y$};
     \node[state,fill=red!10,below of=0Y1] (s2)   {$S,2$};	
    \node[state,right of=s2,fill=green!10,xshift=1cm,] (02)   {$A,2$};
    
    \node[state,right of=02,xshift=1cm,fill=red!10] (0Y2)   {$S,2N$};
    
     \node[right of=0Y2] (4)   {$\cdots$};
    \node[state,above of=4,fill=green!10,xshift=-2cm] (03)   {$A,(D-1)$};
    \node[state,fill=red!10,left of=03,xshift=-1cm] (s3)   {$S,(D-1)$};	
    \node[state,right of=03,xshift=1cm,fill=red!10] (0Y3)   {$S,(D-1)N$};
    
     \node[state,above of=03,fill=green!10] (5)   {$AY,(D-1)Y$};
    \node[state,above of=5,fill=green!10] (04)   {$A,D$};
    \node[state,fill=red!10,left of=04,xshift=-1cm] (s4)   {$S,D$};	
    \node[state,right of=04,xshift=1cm,fill=red!10] (0Y4)   {$S,DN$};


    \path
     
        (0) edge   node[above] {$N$}  (0Y)
        (0Y) edge [bend left]  node[below] {$E$}  (0)
        (0) edge   node[left] {$Y$}  (2)
         (2) edge   node[above] {$0,1$}  (s1)
          (s1) edge node[above] {$E$}  (01)
        (01) edge   node[above] {$N$}  (0Y1)
        (0Y1) edge [bend left]  node[above] {$E$}  (01)
        
         (01) edge   node[left,yshift=-0.2cm] {$Y$}  (3)
         (3) edge   node[above] {$0,1$}  (s2)
          (s2) edge node[above] {$E$}  (02)
        (02) edge   node[above] {$N$}  (0Y2)
        (0Y2) edge [bend left]  node[above] {$E$}  (02)
        
         (02) edge[bend right]   node[above] {$Y$}  (4)
         (4) edge   node[above] {$0,1$}  (s3)
        (s3) edge node[above] {$E$}  (03)
        (03) edge   node[above] {$N$}  (0Y3)
        (0Y3) edge [bend left]  node[below] {$E$}  (03)
        
         (03) edge   node[left] {$Y$}  (5)
         (5) edge   node[above] {$0,1$}  (s4)
        (s4) edge node[above] {$E$}  (04)
        (04) edge   node[above] {$N$}  (0Y4)
        (0Y4) edge [bend left]  node[below] {$E$}  (04);
       
\draw[blue, very thick] (-2.3,0.8) rectangle (4.2,-3.6);			
\end{tikzpicture}
}
\caption{\label{fig:system-intruder-game-attack-D} Game structure between the intruder and the system under bounded $D$ state attacks.}
\end{figure}

Regarding Problem~1, 
it is necessary to construct an observer under a bounded number of state attacks to capture all scenarios in which the intruder might perform state attacks and obtain possible current states of the system based on the obtained attack results as the system evolves.  
Therefore, we need to compose the observer of the system attack model with the game structure to illustrate the possible current states of the system as inferred by the intruder (following the performed state attacks and the obtained attack results as the system evolves).

\begin{mydef}
    \label{def:observation-attack-model}
 (Attack-observer)
Consider an NFA $G=(X,E,\delta,X_{0})$ with all events being observable, under a bounded number of state attacks captured by $(X_{A},D)$, where $X_{A}$ is the set of attacked states and $D \in \mathbb{N}$ is the maximum number of attacks.
Let $G_{siD}=(X_{siD},E_{siD},f_{siD},x_{0,siD})$ be the game-structure automaton under $(X_{A},D)$, $G_{a}$ be the system attack model of $G$ with respect to $X_{A}$, and $Obs(G_{a})$ be the observer of $G_{a}$.
The attack-observer is the composition of $G_{siD}$ and  $Obs(G_{a})$, denoted by $AObs(G)=G_{aobs}=(X_{aobs},E_{aobs},f_{aobs},x_{0,aobs})$.
    $\hfill \square$
\end{mydef}

\begin{myeg}
    \label{eg:composition}
    (Example~\ref{eg:observer} continued) 
    Reconsider the NFA in Fig.~\ref{fig:system} with $X_{0}=\{1,10\}$.  Assume that $D=1$. 
      Following  Definition~\ref{def:number-attack},  the number attack model is shown in Fig.~\ref{fig:number-attack} bounded by the blue line.
     Following Definition~\ref{def:system-intruder-game-attack}, by composing the automaton in Fig.~\ref{fig:system-intruder-game} and the automaton in Fig.~\ref{fig:number-attack}  bounded by the blue line, the game structure with $D=1$ is depicted in Fig.~\ref{fig:system-intruder-game-attack-D}  bounded by the blue line.
    By Definition~\ref{def:observation-attack-model}, by composing the automaton in Fig.~\ref{fig:system-intruder-game-attack-D}  bounded by the blue line and the automaton in Fig.~\ref{fig:observer_system-attack}, the attack-observer under a bounded number of state attacks captured by $(X_{A},D)=(\{2,4\},1)$
 is depicted in Fig.~\ref{fig:composition}, where  
the initial state is $x_{0,aobs}=(A,0,\{1,10\})$.
 $\hfill \triangle$\end{myeg}

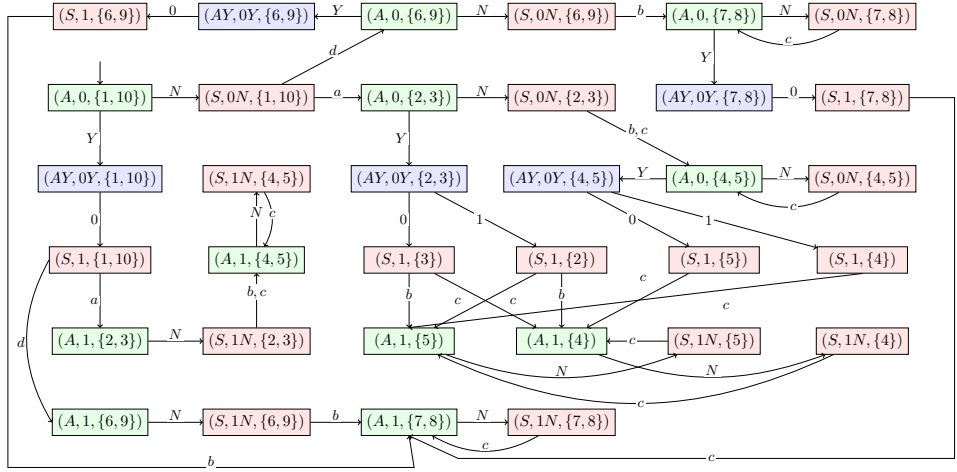
\begin{figure*}[!htbp] 
\centering
\scalebox{0.6}{\begin{tikzpicture}[
  sstate/.style={rectangle, draw, minimum width=2cm, minimum height=0.5cm, fill=red!10, align=center},
  astate/.style={rectangle, draw, minimum width=2cm, minimum height=0.5cm, fill=green!10, align=center},
   aystate/.style={rectangle, draw, minimum width=2cm, minimum height=0.5cm, fill=blue!10, align=center},
  emptycell/.style={draw=none, fill=none, minimum width=2cm, minimum height=0.5cm},
  labelstyle/.style={midway, above, font=\small, fill=white, inner sep=1pt},
  initial text=,
  initial distance=1cm
]

\matrix (m) [matrix of nodes, row sep=1.2cm, column sep=0.8cm, nodes in empty cells] {
  \node[sstate] (s0) {$(S,1,\{6,9\})$}; &
  \node[aystate] (s1) {$(AY,0Y,\{6,9\})$}; &
  \node[astate] (s2) {$(A,0,\{6,9\})$}; &
  \node[sstate] (s3) {$(S,0N,\{6,9\})$}; &
  \node[astate] (s4) {$(A,0,\{7,8\})$}; &
  \node[sstate] (s5) {$(S,0N,\{7,8\})$}; \\
  \node[astate] (s6) {$(A,0,\{1,10\})$}; &
  \node[sstate] (s7) {$(S,0N,\{1,10\})$}; &
  \node[astate] (s8) {$(A,0,\{2,3\})$}; &
  \node[sstate] (s9) {$(S,0N,\{2,3\})$}; &
  \node[aystate] (s10) {$(AY,0Y,\{7,8\})$}; &
  \node[sstate] (s11) {$(S,1,\{7,8\})$}; \\
  \node[aystate] (s12) {$(AY,0Y,\{1,10\})$}; &
  \node[sstate] (s13) {$(S,1N,\{4,5\})$}; &
  \node[aystate] (s14) {$(AY,0Y,\{2,3\})$}; &
  \node[aystate] (s15) {$(AY,0Y,\{4,5\})$}; &
  \node[astate] (s16) {$(A,0,\{4,5\})$}; &
  \node[sstate] (s17) {$(S,0N,\{4,5\})$}; \\
  \node[sstate] (s18) {$(S,1,\{1,10\})$}; &
  \node[astate] (s19) {$(A,1,\{4,5\})$}; &
  \node[sstate] (s20) {$(S,1,\{3\})$}; &
  \node[sstate] (s21) {$(S,1,\{2\})$}; &
  \node[sstate] (s22) {$(S,1,\{5\})$}; &
  \node[sstate] (s23) {$(S,1,\{4\})$}; \\
  \node[astate] (s24) {$(A,1,\{2,3\})$}; &
  \node[sstate] (s25) {$(S,1N,\{2,3\})$}; &
  \node[astate] (s26) {$(A,1,\{5\})$}; &
  \node[astate] (s27) {$(A,1,\{4\})$}; &
  \node[sstate] (s28) {$(S,1N,\{5\})$}; &
  \node[sstate] (s29) {$(S,1N,\{4\})$}; \\
  \node[astate] (s30) {$(A,1,\{6,9\})$}; &
  \node[sstate] (s31) {$(S,1N,\{6,9\})$}; &
  \node[astate] (s32) {$(A,1,\{7,8\})$}; &
  \node[sstate] (s33) {$(S,1N,\{7,8\})$}; &
  \node[emptycell] {}; &
  \node[emptycell] {}; \\
};

\draw[->] (s1) -- node[labelstyle] {$0$} (s0);
\draw[->] (s10) -- node[labelstyle] {$0$} (s11);
\draw[->] (s12) -- node[labelstyle,left] {$0$} (s18);
\draw[->] (s14) -- node[labelstyle,left] {$0$} (s20);
\draw[->] (s15) -- node[labelstyle,left] {$0$} (s22);

\draw[->] (s14) -- node[labelstyle,left] {$1$} (s21);
\draw[->] (s15) -- node[labelstyle,left] {$1$} (s23);

\draw[->] (s2) -- node[labelstyle] {$Y$} (s1);
\draw[->] (s6) -- node[labelstyle,left] {$Y$} (s12);
\draw[->] (s8) -- node[labelstyle,left] {$Y$} (s14);
\draw[->] (s4) -- node[labelstyle,left] {$Y$} (s10);
\draw[->] (s16) -- node[labelstyle,above] {$Y$} (s15);

\draw[->] (s2) -- node[labelstyle] {$N$} (s3);
\draw[->] (s4) -- node[labelstyle] {$N$} (s5);
\draw[->] (s6) -- node[labelstyle] {$N$} (s7);
\draw[->] (s8) -- node[labelstyle] {$N$} (s9);
\draw[->] (s16) -- node[labelstyle] {$N$} (s17);
\draw[->] (s19) -- node[labelstyle] {$N$} (s13);
\draw[->] (s24) -- node[labelstyle] {$N$} (s25);
\draw[->,bend right=20] (s26) to node[labelstyle] {$N$} (s28);
\draw[->,bend right=20] (s27) to node[labelstyle] {$N$} (s29);
\draw[->] (s30) -- node[labelstyle] {$N$} (s31);
\draw[->] (s32) -- node[labelstyle] {$N$} (s33);

\draw[->] (s18) -- node[labelstyle,left] {$a$} (s24);
\draw[->] (s7) -- node[labelstyle] {$a$} (s8);

\draw[->,bend right] (s18.west) to node[labelstyle,left] {$d$} (s30.west);
\draw[->] (s7) -- node[labelstyle] {$d$} (s2);

\draw[->] (s25) -- node[labelstyle] {$b,c$} (s19);
\draw[->] (s9) -- node[labelstyle] {$b,c$} (s16);

\draw[->] (s31) to node[labelstyle] {$b$} (s32);
\draw[->]
  (s0.west)
  -- ++(-1cm, 0) -- ++(0,-10cm) -- ++ (9cm,0) 
  node[labelstyle] {$b$}
  -- (s32.south);
\draw[->] (s3) to node[labelstyle] {$b$} (s4);
\draw[->] (s20) to node[labelstyle] {$b$} (s26);
\draw[->] (s21) to node[labelstyle] {$b$} (s27);

\draw[->,bend left] (s33) to node[labelstyle] {$c$} (s32);
\draw[->,bend left] (s13) to node[labelstyle] {$c$} (s19);
\draw[->] (s20) to node[labelstyle,left,xshift=-0.5cm] {$c$} (s27);
\draw[->] (s21) to node[labelstyle,right,xshift=0.5cm] {$c$} (s26);
\draw[->] (s28) to node[labelstyle,left] {$c$} (s27);
\draw[->,bend left=25] (s29) to node[labelstyle,right] {$c$} (s26);
\draw[->,bend left] (s17) to node[labelstyle,right] {$c$} (s16);
\draw[->,bend left] (s11.east)   -- ++(1cm, 0) -- ++(0,-8cm) -- ++ (-11cm,0)  node[labelstyle,right] {$c$} -- (s32.south);
\draw[->,bend left] (s5) to node[labelstyle,above] {$c$} (s4);
\draw[->] (s22) to node[labelstyle,right,yshift=0.5cm] {$c$} (s27);
\draw[->] (s23.south) to node[labelstyle,below,xshift=2cm] {$c$} (s26.north);
\draw[->] ([yshift=0.5cm]s6.north) -- (s6.north);

\end{tikzpicture}}
\caption{\label{fig:composition}Attack-observer for the NFA in Fig. \ref{fig:system} under a bounded number of state attacks with $X_{A}=\{2,4\}$ and $D=1$.}
\end{figure*}

\subsection{Current-state anonymity (opacity)}

Before formalizing  current-state anonymity (resp. opacity) violation  and attack-enforced anonymity (resp. opacity) violation, we first make the following necessary observations. 
Given a state $x_{aobs}=(x_{aobs}^{1},x_{aobs}^{2},x_{aobs}^{3})\in X_{aobs}$, we denote by $x_{aobs}^{1}$ the situation of the intruder, denote by $x_{aobs}^{2}$ the situation of attacks that the intruder has performed thus far, and denote by $x_{aobs}^{3}$ the state estimate of the system at the intruder.

Let $s=s[0]s[1]s[2]\cdots s[k]\in L(G)$ be a string generated by the system $G$, where $k=|s|\in \mathbb{N}$, $s[0]=\varepsilon$, and $s[j]\in E$ for all $j\in \{1,2, \cdots,k\}$.
We denote by $r_{a}[i]\in \{Y,N\}$ and $r_{ar}[i]\in \{0,1,\varepsilon\}$ the attack action performed by the intruder and the attack result obtained by the intruder after observing $s[i]$, respectively, where $i\in \{0,1,2,\cdots,k\}$. 
If the intruder launches a state attack, i.e., $r_{a}[i]=Y$, then the attack result will be either 0 or 1;
if the intruder does not launch a state attack, i.e., $r_{a}[i]=N$, then there is no attack result, i.e., the attack result is $\varepsilon$. 
Therefore, we have $r_{a}[i]r_{ar}[i]\in \{Y0,Y1,N\}$.
Following the game-structure automaton in Fig.~\ref{fig:system-intruder-game-attack-D}, the full attack trace that is available to the intruder is denoted by $s_{a}=s[0]r_{a}[0]r_{ar}[0]s[1]r_{a}[1]r_{ar}[1]s[2]r_{a}[2]r_{ar}[2]\cdots s[k]r_{a}[k]r_{ar}[k]$, where
$r_{a}=r_{a}[0]r_{a}[1]r_{a}[2]\cdots r_{a}[k]$ and $r_{ar}=r_{ar}[0]r_{ar}[1]$ $r_{ar}[2]\cdots r_{ar}[k]$ are the sequence of attack actions performed by the intruder and the sequence of attack results obtained by the intruder associated with $s$, respectively.
The number of state attacks that have been performed is denoted by $|r_{a}|_{Y}$.
Note that since the maximum number of state attacks is bounded by $D$, we have  $|r_{a}|_{Y}\le D$; this can be guaranteed by the composition of the game structure between the intruder and the system and the number attack model.

{
\begin{lem}\label{lem:attack-observer}
  Consider an NFA $G=(X,E,\delta,X_{0})$ with all events being observable, under a bounded number of state attacks captured by $(X_{A},D)$, where $X_{A}$ is the set of attacked states and $D \in \mathbb{N}$ is the maximum number of attacks.
    The set of attack actions is denoted by $E_{A}=\{Y,N\}$, the set of attack results is denoted by $E_{ar}=\{0,1,\varepsilon\}$, the set of possible concatenations is denoted by  $E_{aar}=\{Y0,Y1,N\}$, and
 the corresponding attack-observer is denoted by $G_{aobs}=(X_{aobs},E_{aobs},f_{aobs},x_{0,aobs})$. The attack-observer $G_{aobs}$ captures all $s\in L(G)$ and all possible $r_{a}$ for $s$.

Proof:    Due to Definition~\ref{def:observation-attack-model}, the observation model $G_{aobs}$ is the composition of the game structure $G_{siD}$ under $(X_{A},D)$ and the observer of the system attack model $Obs(G_{a})$.
Based on Definition~\ref{def:system-intruder-game-attack}, $G_{siD}$ is the composition of the game structure between the intruder and the system $G_{si}$ and the number attack model $G_{D}$.
Based on Definitions~\ref{def:game-structure-automaton},  \ref{def:number-attack}, and \ref{def:system-attack}, $G_{si}$ captures the interaction between the system and the intruder when there exists state attack $X_{A}$,  $G_{D} $ captures the intruder attack ability for a maximum number of $D$ state attacks, and $G_{a}$ captures 
the current state of the system as the system evolves and the attack results are obtained. 
Therefore, $G_{aobs}$ captures all $s\in L(G)$ and all possible $r_{a}$ for $s$.
 
\end{lem}
}

\begin{mydef}\label{def:anonymity-string}
(Anonymity-violating attack sequence under a bounded number of state attacks)
   Consider an NFA $G=(X,E,\delta,X_{0})$ with all events being observable, under a bounded number of state attacks captured by $(X_{A},D)$, where $X_{A}$ is the set of attacked states and $D \in \mathbb{N}$ is the maximum number of attacks.
    The set of attack actions is denoted by $E_{A}=\{Y,N\}$, the set of attack results is denoted by $E_{ar}=\{0,1,\varepsilon\}$, the set of possible concatenations is denoted by  $E_{aar}=\{Y0,Y1,N\}$, and
 the corresponding attack-observer is denoted by $G_{aobs}=(X_{aobs},E_{aobs},f_{aobs},x_{0,aobs})$.
 Given a string $s=s[0]s[1]s[2]\cdots s[k]\in L(G)$ with $k=|s|$, an attack sequence $r_{a}=r_{a}[0]r_{a}[1]r_{a}[2]\cdots r_{a}[k]$ with $r_{a}[i]\in E_{A}$, $i=0,1,\cdots,k$, is said to be anonymity-violating under $(X_{A},D)$ for $s$, if  
   \[(\exists r_{ar}=r_{ar}[0]r_{ar}[1]r_{ar}[2]\cdots r_{ar}[k-1],\forall r_{ar}[k]\in E_{ar} )\] \[[r_{a}[i]r_{ar}[i]\in E_{aar} 
   ~\wedge~ x_{aobs}=f_{aobs}(x_{0,aobs},s_{a})\neq \emptyset\implies\] \[|x_{aobs}^{3}|=1]\]
where $|r_{a}|_{Y}\le D $, $s_{a}=s[0]r_{a}[0]r_{ar}[0]s[1] r_{a}[1]r_{ar}[1]s[2]$ $r_{a}[2]r_{ar}[2]\cdots s[k]r_{a}[k]r_{ar}[k]$, and $x_{aobs}^{3}$ is the set of current states of the system associated with $x_{aobs}$ (inferred by the intruder).
$\hfill \square$
\end{mydef}

According to Definition~\ref{def:anonymity-string}, a sequence of attack actions $r_a$ is considered an anonymity-violating attack sequence for a trace $s$ if, there exists at least one specific realization of attack actions $r_{a}[0]r_{a}[1]\cdots r_{a}[|s|-1]$ for $s[0]s[1]s[2]\cdots s[|s|-1]$ such that for all possible realizations of the attack $r_{a}[|s|]$ for $s[|s|]$, the resulting full attack trace reaches a state in the observation model in which the current state of the system is uniquely identified.

\begin{myeg}
    \label{eg:current-state-anonymity-violating-string} (Example~\ref{eg:composition} continued)
Reconsider the NFA in Fig.~\ref{fig:system} with $X_{A}=\{2,4\}$ and $D=1$. The corresponding observation model is shown in Fig.~\ref{fig:composition}. Suppose that $s=s[0]s[1]=\varepsilon a$ and $r_{a}=r_{a}[0]r_{a}[1]=NY$.
For $r_{a}[1]=Y$, we have $r_{ar}[1]=0$ (resp. $r_{ar}[1]=1$) satisfying $r_{a}[1]r_{ar}[1]=Y0$ (resp. $r_{a}[1]r_{ar}[1]=Y1$) such that $s_{a}=NaY0$ (resp. $s_{a}=NaY1$), $x_{aobs}=f_{aobs}(x_{0,aobs}$, $s_{a})=f_{aobs}((A,0,\{1,10\}),NaY0)=(S,1,\{3\})$ (resp. $x_{aobs}=f_{aobs}(x_{0,aobs},s_{a})=f_{aobs}((A,0,\{1,10\}),NaY1)=(S,1,\{2\})$, and $|x_{aobs}^{3}|=|\{3\}|=1$ (resp. $|x_{aobs}^{3}|=|\{2\}|=1$).
According to Definition~\ref{def:anonymity-string}, $r_{a}=NY$ with $|r_{a}|_{Y}=1$  is an anonymity-violating attack sequence for $s=a$.
    $\hfill \triangle$
\end{myeg}

\begin{mydef}\label{def:anonymity-system}
{(Current-state anonymity violation under a bounded number of state attacks)}
   Consider an NFA $G=(X,E,\delta,X_{0})$ with all events being observable, under a bounded number of state attacks captured by $(X_{A},D)$, where $X_{A}$ is the set of attacked states and $D \in \mathbb{N}$ is the maximum number of attacks.
   System $G$ is said to be current-state anonymity violating under $(X_{A},D)$ if there exists $s\in L(G)$ for which there exists an anonymity-violating attack sequence $r_{a}$ under $(X_{A},D)$.
$\hfill \square$
\end{mydef}
}

\begin{mydef}
    \label{def:anonymity-attackable-string}
   (Attack-enforced anonymity-violating  attack sequence under a bounded number of state attacks)
     Consider an NFA $G=(X,E,\delta,X_{0})$ with all events being observable, under a bounded number of state attacks captured by $(X_{A},D)$, where $X_{A}$ is the set of attacked states and $D \in \mathbb{N}$ is the maximum number of attacks.
Given  $s=s[0]s[1]s[2]\cdots s[k]\in L(G)$,  an attack sequence $r_{a}=r_{a}[0]r_{a}[1]r_{a}[2]\cdots r_{a}[k]$ is said to be attack-enforced anonymity-violating  for $s$, if for all  
$e\in E$ with $se\in L(G)$, there exist $r_{a}''$ for $e$ and
 $s'=s'[1]$ $s'[2]\cdots s'[p]\in SX(se)$ such that there exists an attack sequence $r_{a}'=r_{a}'[1]r_{a}'[2]\cdots r_{a}'[p]$ for $s'$ satisfying that 
(i) $r_{a}r_{a}''r_{a}'$ is an anonymity-violating attack sequence for $ses'$, and
(ii)  $|r_{a}r_{a}''r_{a}'|_{Y}\le D$,
 where $k=|s|$ and $p=|s'|$.
    $\hfill \square$
\end{mydef}

Definition~\ref{def:anonymity-attackable-string} indicates that an attack sequence $r_{a}$ is attack-enforced anonymity-violating  for $s$, if for all possible subsequent activity $e$ that can be generated by the system, there exists an attack sequence $r_{a}''$ for $e$ and there exists $s'\in SX(se)$ such that there exists an attack sequence $r_{a}'$ for $s'$ satisfying that  $r_{a}r_{a}''r_{a}'$ is an anonymity-violating attack sequence  for $ses'$  and the total number of attacks is less than or equal to $D$.
That also implies that a feasible attack sequence for the aim of attack-enforced current-state anonymity violation should guarantee that an anonymity-violating state will be eventually reached for all possible subsequent activity generated by the system via appropriate attack operations in the future while keeping the total number of state attacks below $D$.

\begin{myeg}
    \label{eg:current-state-anonymity-feasible-attack}
(Example~\ref{eg:current-state-anonymity-violating-string} 
 continued)    Let us consider $s=s[0]s[1]=\varepsilon a$ and $r_{a}=NY$ for $s$.
 We have $e\in \{b,c\}$ satisfying that $se\in L(G)$.
 For $e=b$, we can find $r_{a}''=N$ and $s'=c$ such that there exists $r_{a}'=N$ for $s'=c$ satisfying that $r_{a}r_{a}''r_{a}'=NYNN$ is an anonymity-violating  attack sequence for $ses'=\varepsilon a bc$
based on Definition~\ref{def:anonymity-string}.
To see this, notice that $|r_{a}r_{a}''r_{a}'|_{Y}=1\le D=1$ and in Fig.~\ref{fig:composition}, $f_{aobs}((A,0,\{1,10\}),NaY0bNcN)=(S,1N,\{4\})$ and $f_{aobs}((A,0,\{1,10\}),NaY1bNcN)=(S,1N,\{5\})$. 
We can also find a solution for $e=c$. 
Hence, from Definition~\ref{def:anonymity-attackable-string}, $r_{a}=NY$ is an attack-enforced anonymity-violating attack sequence  for $s=a$.
$\hfill \triangle$
\end{myeg}

\begin{mydef}\label{def:anonymity-attackable}
(Attack-enforced current-state anonymity violation under a bounded number of state attacks)
Consider an NFA $G=(X,E,\delta,X_{0})$ with all events being observable, under a bounded number of state attacks captured by $(X_{A},D)$, where $X_{A}$ is the set of attacked states and $D \in \mathbb{N}$ is the maximum number of attacks.
  System $G$ is said to be attack-enforced anonymity-violating under $(X_{A},D)$ if  for all $s\in L(G)$, the following conditions hold:
\begin{enumerate}
    \item   there exists an attack-enforced anonymity-violating attack sequence $r_{a}=r_{a}[0]r_{a}[1]\cdots$ $r_{a}[|s|]$ for $s$ under the bounded number of state attacks, and
    \item for all $s'\in \overline{s}$,
 $r_{a}'=r_{a}[0]r_{a}[1]\cdots r_{a}[|s'|]$ is an attack-enforced anonymity-violating attack sequence for $s'$ under the bounded number of state attacks.
\end{enumerate}
$\hfill \square$
\end{mydef}

Definition~\ref{def:anonymity-attackable} states that if for all observed strings $s$ that can be generated by the system, we can find an attack-enforced anonymity-violating attack sequence $r_{a}$, and for each prefix of the observed string $s'$, an attack-enforced anonymity-violating sequence $r_{a}'$ can be found in  $r_{a}$, then the system is attack-enforced anonymity-violating. In other words, the system can be forced to an anonymity-violating state regardless of the system activities following the already performed state attacks, the obtained attack results, and the future attack actions and possible attack results.

\section{Verification of current-state anonymity violation under a bounded number of state attacks}
{
{
This section concentrates on the construction of a verifier based on the observation model for  verifying current-state anonymity violation under a bounded number of state attacks. 

Let us consider an observation model $G_{aobs}=(X_{aobs}$, $E_{aobs},\delta_{aobs},x_{0,aobs})$.
Given a state of the observation model $x_{aobs}=(x_{aobs}^{1},x_{aobs}^{2},x_{aobs}^{3})$, similar to the state in $G_{si}$,  $x_{aobs}^{1}$ within $x_{aobs}\in X_{aobs}$
has the following three scenarios:
(i) $x_{aobs}^{1}=S$ represents that it is the system's turn to generate subsequent activities and the intruder's turn to estimate the current state of the system after the previous state attack, 
(ii) $x_{aobs}^{1}=AY$ represents that it is the intruder's turn to obtain the attack result, and 
(iii) $x_{aobs}^{1}=A$ represents that it is the intruder's turn to decide whether to perform a state attack or not.
Based on the characteristic of $x_{aobs}^{1}$, the set of states in the observation model can be partitioned into three types of states:
$X_{aobs}=X_{aobs1}~\dot{\cup}~X_{aobs2}~\dot{\cup}~X_{aobs3}$, where $X_{aobs1}= \{x_{aobs}\mid x_{aobs}^{1}=S \}$ is the set of type-I states, $X_{aobs2}= \{x_{aobs}\mid x_{aobs}^{1}=AY\}$ is the set of type-II states, and $X_{aobs3}= \{x_{aobs}\mid x_{aobs}^{1}=A\}$ is the set of type-III states.

\begin{myeg}
    \label{eg:composition-state-partition} (Example~\ref{eg:composition} continued)
Reconsider the observation model in Fig.~\ref{fig:composition}. 
Each type-I state is colored with red,  each type-II state  is colored with blue, and each type-III state is colored with green.
    $\hfill \triangle$
\end{myeg}

In Definitions~\ref{def:anonymity-violation-state}--\ref{def:intermediate-anonymity-violation-state-1} below, we  consider an NFA $G=(X,E,\delta,X_{0})$ with all events being observable, under a bounded number of state attacks captured by $(X_{A},D)$, where $X_{A}$ is the set of attacked states and $D \in \mathbb{N}$ is the maximum number of attacks.
 Let $G_{aobs}=(X_{aobs}, E_{aobs}, f_{aobs}, x_{0,aobs})$ be the corresponding observation model.

\begin{mydef}
    \label{def:anonymity-violation-state}
    (Anonymity-violating state in an observation model)
 A type-I state $x_{aobs}=(x_{aobs}^{1},x_{aobs}^{2},x_{aobs}^{3})\in X_{aobs1}$ is said to be anonymity-violating if 
     \[|x_{aobs}^{3}|=1.\]
     The set of all anonymity-violating states is denoted by $X_{aobs1}^{av}$.
     $\hfill \square$
\end{mydef}

A type-I state that is anonymity-violating is also said to be an intermediate anonymity-violating state.
We use $X_{aobs1}^{iav}$ to denote the set of intermediate anonymity-violating states.

\begin{mydef}
    \label{def:intermediate-anonymity-violation-state-2}
    (Intermediate anonymity-violating type-II state in an observation model)
 Given the set of intermediate anonymity-violating (type-I) states $X_{aobs1}^{iav}$, a
 type-II state $x_{aobs}=(x_{aobs}^{1},x_{aobs}^{2},x_{aobs}^{3})\in  X_{aobs2}$ is said to be intermediate anonymity-violating if  
  \[(\forall \sigma\in \{0,1\})[f_{aobs}(x_{aobs},\sigma)\neq \emptyset \implies\] \[  [x_{aobs}'=f_{aobs}(x_{aobs},\sigma)\in  X_{aobs1}^{iav}]].\]
 The set of intermediate anonymity-violating type-II states is denoted by $X_{aobs2}^{iav}$.
     $\hfill \square$
\end{mydef}

\begin{mydef}
    \label{def:intermediate-anonymity-violation-state-3}
    (Intermediate anonymity-violating type-III state in an observation model)
 Given the set of intermediate anonymity-violating states $X_{aobs1}^{iav}\cup X_{aobs2}^{iav}$,
a type-III state $x_{aobs}=(x_{aobs}^{1},x_{aobs}^{2},x_{aobs}^{3})\in X_{aobs3}$ is said to be intermediate anonymity-violating if
\[(\exists \sigma\in \{Y,N\})[x_{aobs}'=f_{aobs}(x_{aobs},\sigma)\in  (X_{aobs1}^{iav}\cup X_{aobs2}^{iav})].\]

 The set of intermediate anonymity-violating type-III states is denoted by $X_{aobs3}^{iav}$.
     $\hfill \square$
\end{mydef}

\begin{mydef}
    \label{def:intermediate-anonymity-violation-state-1}
    (Intermediate anonymity-violating type-I state in an observation model)
  Given the set of intermediate anonymity-violating type-III states $X_{aobs3}^{iav}$,
  a type-I state $x_{aobs}=(x_{aobs}^{1},x_{aobs}^{2},x_{aobs}^{3})\in X_{aobs1}$ is said to be intermediate anonymity-violating if  
  \[(\exists e\in E)[x_{aobs}'=f_{aobs}(x_{aobs},e)\in  X_{aobs3}^{iav}].\]
     $\hfill \square$
\end{mydef}

With the initial set of intermediate anonymity-violating type-I states,
we can compute the set of intermediate anonymity-violating type-II states according to Definition~\ref{def:intermediate-anonymity-violation-state-2}. 
Following Definition~\ref{def:intermediate-anonymity-violation-state-3},  the set of intermediate anonymity-violating type-III states  can be obtained.
Then, by Definition~\ref{def:intermediate-anonymity-violation-state-1}, new intermediate anonymity-violating type-I states can be obtained.
Hence, the new set of  intermediate anonymity-violating type-I states is obtained following Definition~\ref{def:intermediate-anonymity-violation-state-1}.
By iteratively repeating the construction of new  intermediate anonymity-violating type-I states by Definition~\ref{def:intermediate-anonymity-violation-state-1}, the construction of new  anonymity-violating type-II states by Definition~\ref{def:intermediate-anonymity-violation-state-2}, and the construction of new  anonymity-violating  type-III states by Definition~\ref{def:intermediate-anonymity-violation-state-3},  until no new intermediate anonymity-violating type-I state is generated, we can obtain the final set of all intermediate anonymity-violating states, denoted by $X_{aobs}^{iav}=X_{aobs1}^{iav}\cup X_{aobs2}^{iav}\cup X_{aobs3}^{iav}$.
The process is detailed in Algorithm~\ref{alg:anonymity-violating-states} below.

\begin{algorithm}[!htbp]
    \caption{Obtaining the set of all intermediate anonymity-violating states for a given observation model under a bounded number of state attacks.}\label{alg:anonymity-violating-states}
    \LinesNumbered
    \KwIn{ Observation model $G_{aobs}=(X_{aobs},E_{aobs},f_{aobs},x_{0,aobs})$ for an NFA  $G=(X,E,\delta,X_{0})$ under a bounded number of state attacks $(X_{A},D)$. }
    \KwOut{ Set of all intermediate anonymity-violating states $X_{aobs}^{iav}$.}
     Initialize $X_{aobs1}^{iav}=X_{aobs2}^{iav}=X_{aobs3}^{iav}=\emptyset$\;
    Compute the set of anonymity-violating type-I states $X_{aobs1}^{av}$ by Definition~\ref{def:anonymity-violation-state}\;
    Set $X_{aobs1}^{iav}=X_{aobs1}^{av}$ be the starting set of intermediate anonymity-violating type-I states\;
    Compute the set of intermediate anonymity-violating type-II states $X_{aobs2}^{iav}$ by Definition~\ref{def:intermediate-anonymity-violation-state-2}\;
    Obtain the set of intermediate anonymity-violating type-III states $X_{aobs3}^{iav}$ based on  $X_{aobs1}^{iav}$ and $X_{aobs2}^{iav}$  by Definition~\ref{def:intermediate-anonymity-violation-state-3}\;
     \For{ $x_{aobs3}\in X_{aobs3}\setminus X_{aobs3}^{iav}$ that has not been examined}{  
    $ X{_{aobs1}^{iav}}'=\emptyset$\;
      \For{ $x_{aobs1}\in X_{aobs1}\setminus X_{aobs1}^{iav}$ that has not been examined}{
  \If{$x_{aobs1}$ is an intermediate anonymity-violating type-I state based on Definition~\ref{def:intermediate-anonymity-violation-state-1}}{
  Add $x_{aobs1}$ to $X{_{aobs1}^{iav}}'$\;
  }
  Mark $x_{aobs1}$ as examined\;
  }
   Mark $x_{aobs3}$ as examined\;
}
\If{$X{_{aobs1}^{iav}}'=\emptyset$}{
 Return $X_{aobs}^{iav}=X_{aobs1}^{iav}\cup X_{aobs2}^{iav}\cup X_{aobs3}^{iav}$.
}
\Else{
$X_{aobs1}^{iav}=X_{aobs1}^{iav}\cup X{_{aobs1}^{iav}}'$\;
Compute the set of intermediate anonymity-violating type-II states $X_{aobs2}^{iav}$ by Definition~\ref{def:intermediate-anonymity-violation-state-2}\;
    Obtain the set of intermediate anonymity-violating type-III states $X_{aobs3}^{iav}$ by Definition~\ref{def:intermediate-anonymity-violation-state-3}\;
    Go to Line~6.
  }
  
\end{algorithm}

\begin{mycom}
    Complexity analysis of Algorithm~\ref{alg:anonymity-violating-states} is provided in what follows.
In the worst-case scenario, for an observation model, the number of type-I states is $(2D+1)2^{|X|}$, the number of type-II states is $(D+1)2^{|X|}$, and the number of type-III states is $D2^{|X|}$.
For each type-I state, there are at most $|E|$ edges, for each type-II state, there are at most $2$ edges, and for each type-III state, there are at most 2 edges.
Therefore, the number of edges in the observation model is at most $(2D|E|+4D+|E|+2)2^{|X|}$.
From Algorithm~\ref{alg:anonymity-violating-states}, each state should be checked whether it is an intermediate anonymity-violating state. 
Notice that although we need to iteratively compute new intermediate anonymity-violating type-I state, once a type-I state is examined, we do not need to examine it again.
In conclusion, the complexity of obtaining all intermediate anonymity-violating states based on the observation model  is $O((2D|E|+4D+|E|+2)2^{|X|})$.

  \end{mycom}

\begin{myeg}
    \label{eg:anonymity-violating-state}
 (Example~\ref{eg:composition-state-partition} continued)   Reconsider the observation model in Fig.~\ref{fig:composition}. 
According to Definition~\ref{def:anonymity-violation-state}, the set of anonymity-violating states is $X_{aobs1}^{av}=$ $\{(S,1,\{2\})$, $(S,1,\{3\})$, $
(S,1,\{4\})$, $(S,1,\{5\})$, $(S,1N,\{4\})$, $(S,1N,\{5\}) \}$.
Following Line~3, the starting set of intermediate anonymity-violating type-I states is  $X_{aobs1}^{iav}=$ $\{(S,1,\{2\})$, $(S,1,\{3\})$, $
(S,1,\{4\})$, $(S,1,\{5\})$, $(S,1N,\{4\})$, $(S,1N,\{5\}) \}$.
In Line~4, based on Definition~\ref{def:intermediate-anonymity-violation-state-2}, the set of intermediate anonymity-violating type-II states is 
$X_{aobs2}^{iav}=\{(AY,0Y,\{2,3\})$, $(AY,0Y,\{4,5\})\}$.
In Line~5, by Definition~\ref{def:intermediate-anonymity-violation-state-3}, the set of intermediate anonymity-violating type-III states is obtained as $X_{aobs3}^{iav}=\{(A,0,\{2,3\})$, $(A,0,\{4,5\})$, $(A,1,\{4\})$, $(A,1,\{5\})\}$.
From lines~6--12, according to Definition~\ref{def:intermediate-anonymity-violation-state-1}, the new set of  intermediate anonymity-violating type-I states is $X{_{aobs1}^{iav}}'=$ $\{ (S,0N,\{1,10\})$, $(S,0N,\{2,3\})$, $(S,0N,\{4,5\})\}$.
Then, following lines~16--19, 
$X{_{aobs1}^{iav}}=$ $\{(S,1,\{2\})$, $(S,1,\{3\})$, $
(S,1,\{4\})$, $(S,1,\{5\})$, $(S,1N,\{4\})$, $(S,1N,\{5\})$, $ (S,0N,\{1,10\})$, $(S,0N,\{2,3\})$, $(S,0N,\{4,5\})\}$;
based on Definition~\ref{def:intermediate-anonymity-violation-state-2}, the new set of intermediate anonymity-violating type-II states is $X_{aobs2}^{iav}=\{(AY,0Y,\{2,3\})$, $(AY,0Y,\{4,5\})\}$;  by Definition~\ref{def:intermediate-anonymity-violation-state-3}, the new set of intermediate anonymity-violating type-III states is $X_{aobs3}^{iav}=\{(A,0,\{2,3\})$, $(A,0,\{4,5\})$, $(A,1,\{4\})$, $(A,1,\{5\})$, $(A,0,\{1,10\})\}$.

Then, we go to lines 6--12, where the new set of intermediate anonymity-violating type-I states is  $X{_{aobs1}^{iav}}'=\emptyset$.
Since there does not exist $x_{aobs3}\in X_{aobs3}^{iav}$ that has not been examined, the set of  all intermediate anonymity-violating states is obtained, denoted by $X_{aobs}^{iav}=$ $X_{aobs1}^{iav}\cup X_{aobs2}^{iav} \cup X_{aobs3}^{iav}=\{(S,1,\{2\})$, $(S,1,\{3\})$, $
(S,1,\{4\})$, $(S,1,\{5\})$, $(S,1N,\{4\})$, $(S,1N,\{5\})$, $(S,0N$, $\{1,10\})$, $(S,0N,\{2,3\})$, $(S,0N,\{4,5\})$, $(AY,0Y,\{2,3\})$, $(AY,0Y,\{4,5\})$, $(A,0,\{2,3\})$, $(A,0,\{4,5\})$, $(A,1,\{4\})$, $(A,1,\{5\})$, $(A,0,\{1,10\})\}$
   $\hfill \triangle$
\end{myeg}

\begin{mydef}
    \label{def:verifier}
    Consider an NFA $G=(X,E,\delta,X_{0})$ with all events being observable, under a bounded number of state attacks captured by $(X_{A},D)$, where $X_{A}$ is the set of attacked states and $D \in \mathbb{N}$ is the maximum number of attacks.
    Let $G_{aobs}=(X_{aobs}, E_{aobs}, f_{aobs}, x_{0,aobs})$ be the corresponding observation model, and  $X_{aobs}^{iav}$ be the set of all intermediate anonymity-violating states.
    The verifier is denoted by $G_{v}=(X_{v}, E_{v}, f_{v}, x_{0,v})$, where 
    \begin{enumerate}
        \item $X_{v}\subseteq X_{aobs}^{iav}$ is the set of states;
        \item $E_{v}=E\cup\{Y,N\}\cup \{0,1\}$ is the set of events;
        \item $x_{0,v}=x_{0,aobs}$ is the initial state;
        \item $f_{aobs}$ is the transition function, defined as:\\ for all $x_{v}\in X_{v}$ and for all $\sigma\in E_{v}$, $f_{v}(x_{v},\sigma)=f_{aobs}(x_{v},\sigma)$, if there exists $x_{aobs}'\in X_{aobs}^{iav}$ such that $x_{aobs}'=f_{aobs}(x_{v},\sigma)$, otherwise $f_{v}(x_{v},\sigma)$ is not defined.
    \end{enumerate}
    $\hfill \square$
\end{mydef}

Algorithm~\ref{alg:verification-algorithm} below focuses on the verification of an anonymity-violating system under a bounded number of state attacks.

\begin{algorithm}[!htbp]
    \caption{Verification of current-state anonymity violation under a bounded number of state attacks.}\label{alg:verification-algorithm}
    \LinesNumbered
    \KwIn{An NFA  $G=(X,E,\delta,X_{0})$ under a bounded number of state attacks $(X_{A},D)$. }
    \KwOut{NFA violates current-state anonymity under $(X_{A},D)$ or not.}
   Obtain the system attack model  with respect to  $X_{A}$ by Definition~\ref{def:system-attack}: $G_{a}=(X_{a}, E_{a},\delta_{a},X_{0,a})$\;
   Construct the number attack model with respect to $D$ by Definition~\ref{def:number-attack}: $G_{D}=(X_{D},E_{D},\delta_{D},X_{0,D})$\;
    Obtain the  game-structure automaton by Definition~\ref{def:game-structure-automaton}: $G_{si}=(X_{si},E_{si},f_{si},x_{0,si})$\;
   Construct the game-structure automaton with $D$ by Definition~\ref{def:system-intruder-game-attack}: $G_{siD}=(X_{siD},E_{siD},f_{siD},x_{0,siD})$\;

   Acquire the observer of 
$G_{a}$ by Definition~\ref{def:observer}: $Obs(G_{a})=(X_{a,obs},E_{a,obs},f_{a,obs},x_{0,a,obs})$\;
   Obtain the observation model by composing $G_{siD}$ and  $Obs(G_{a})$ by Definition~\ref{def:compostion}: $G_{aobs}=(X_{aobs},E_{aobs},f_{aobs},x_{0,aobs})$\;
   Construct the verifier of $G_{aobs}$ by Definition~\ref{def:verifier}: $G_{v}=(X_{v},E_{v},f_{v},x_{0,v})$;
  
\If{$G_{v}$ is not an empty automaton }
                {NFA violates  current-state anonymity under $(X_{A},D)$. }
\Else{NFA satisfies current-state anonymity under $(X_{A},D)$.}

\end{algorithm}

\begin{mycom}
    The complexity of verification for current-state anonymity violation under  a bounded number of state attacks is analyzed.
The system attack model $G_{a}$ has at most $|X|$ states and $|X|(|E|+1)$ edges.
The number attack model has at most $3D+2$ states and $(2D+1)|E|+4D+1$ edges.
The game-structure automaton between the intruder and the system has at most $3$ states and $4+|E|$ edges.
The game structure with $D$ has at most $4D+2$ states and $(2D+1)|E|+4D+2$ edges.
The observer of $G_{a}$ has at most $2^{|X|}$ states and $(|E|+2)2^{|X|}$.
The observation model 
has at most $(4D+2)2^{|X|}$ states and $((2D+1)|E|+4D+2)2^{|X|}$ edges.
The set of  all intermediate anonymity-violating states can be obtained with complexity of
$((2D+1)|E|+4D+2)2^{|X|}$.
In the worst case, the verifier is constructed with complexity $((2D+1)|E|+4D+2)2^{|X|}$.
In the end, verification for current-state anonymity violation under a bounded number of state attacks can be done with complexity $O(((2D+1)|E|+4D+2)2^{|X|})$.
  \end{mycom}

\begin{thm}
    \label{thm:anonymity-violation}
     (Verification of current-state anonymity violation for a system under a bounded number of state attacks)
 Consider an NFA $G=(X,E,\delta,X_{0})$ with all events being observable, under a bounded number of state attacks captured by $(X_{A},D)$, where $X_{A}$ is the set of attacked states and $D \in \mathbb{N}$ is the maximum number of attacks.
 Let  $G_{aobs}=(X_{aobs},E_{aobs},f_{aobs},x_{0,aobs})$ and $G_{v}=(X_{v}, E_{v}, f_{v}, x_{0,v})$ be the corresponding attack-observer and $(X_{A},D)$-verifier, respectively.
System $G$ violates current-state anonymity under $(X_{A},D)$ if  and only if $G_{v}$ is not an empty automaton.  
 
 {Proof:
(If) We need to show that if  $G_{v}$ is not an empty automaton, the system $G$ violates current-state anonymity under $(X_{A},D)$.

Based on the construction of
$G_{v}$ by Definition~\ref{def:verifier}, we know that all $x_{v}\in X_{v}$ is intermediate anonymity-violating.
Following Algorithm~\ref{alg:anonymity-violating-states} and Definition~\ref{def:anonymity-violation-state}, there necessarily exists the starting set of anonymity-violating states, denoted by $X_{v}^{av}\subset X_{aobs1}^{av}$.
Let us consider an arbitrary anonymity-violating state $x_{v}\in X_{v}^{av}$, i.e., $|x_{v}^{3}|=1$.
For $x_{v}$, the following two scenarios should be considered: (1) $x_{v}^{2}\in \{0N,1N,\cdots,DN\}$ and (2) $x_{v}^{2}\in \{0,1,\cdots,D\}$.
For scenario (1), by Definition~\ref{def:intermediate-anonymity-violation-state-3}, we can directly find an intermediate anonymity-violating type-III state  $x_{v}'\in X_{v}$ and $r_{a}r_{ar}=N$ such that $x_{v}'=f_{v}(x_{v},N)$.
For scenario (2), we can first find 
 an intermediate anonymity-violating type-II state $x_{v}''\in X_{v}$ based on Definition~\ref{def:intermediate-anonymity-violation-state-2}, satisfying that for all $r_{ar}\in \{0,1\}$ with $f_{v}(x_{v}'',\sigma)\neq \emptyset$ such that $x_{v}=f_{v}(x_{v}'',\sigma)$, and then, by Definition~\ref{def:intermediate-anonymity-violation-state-3},  we can find an intermediate anonymity-violating type-III state  $x_{v}'''\in X_{v}$ and $r_{a}=Y$ such that $x_{v}'''=f_{v}(x_{v}'',Y)$.
 For both scenarios, we can find  an anonymity-violating type-III state.
 Following Definition~\ref{def:intermediate-anonymity-violation-state-1}, we can find a new (different) intermediate anonymity-violating type-I state $x_{v}''''$ and an event $e\in E$ such that $x_{v}=f_{v}(x_{v}'''',er_{a}r_{ar})$, where $r_{a}r_{ar}\in \{Y0,Y1,N\}$.

 Note that the number of type-I states in $G_{v}$ and $G_{aobs}$ is bounded by $2^{|X|}(2D+1)$.
 Assume that $n< 2^{|X|}(2D+1) \in \mathbb{N}^{+}$ is an arbitrary non-negative integer.
 Let $x_{v}=x_{n,v}$, $x_{v}''''=x_{n-1,v}$, $e=e_{n}$, $r_{a}=r_{n,a}$, and $r_{ar}=r_{n,ar}$.
 By iteratively finding new (different) intermediate anonymity-violating type-I states for $n-2$ steps, we can find $x_{1,v}$ such that one of the following condition holds: (1) $x_{1,v}^{3}=x_{0,v}^{3}$ and $x_{1,v}^{2}=0N$; (2) $x_{1,v}^{3}\subseteq x_{0,v}^{3}$ and $x_{1,v}^{2}=1$.
 For scenario (1), we can find $e_{0}=\varepsilon$, $r_{0,a}=N$, and $r_{0,ar}=\varepsilon$ such that $x_{1,v}=f_{v}(x_{0,v},e_{0}r_{0,a}r_{0,ar})$.
  For scenario (2), we can find $e_{0}=\varepsilon$, $r_{0,a}=Y$, and $r_{0,ar}\in \{0,1\}$ such that $x_{1,v}=f_{v}(x_{0,v},e_{0}r_{0,a}r_{0,ar})$.

 Letting $e_{0}e_{1}e_{2}\cdots e_{n}=s$ and $n=|s|$,
we can find a string $s_{a}'=s[0]r_{a}[0]r_{ar}[0]s[1]r_{a}[1]r_{ar}[1]s[2]r_{a}[2]r_{ar}[2]\cdots$ $s[|s|-1] r_{a}[|s|-1]r_{ar}[|s|-1]\in L(G_{v})$ such that (1) $f_{v}(x_{0,v},s_{a}')=x_{v}'$ and  for  $i\in \{0,1,2,\cdots,|s|-1\}$,  $s[i]\in E\cup \{\varepsilon\}$ and $r_{a}[i]r_{ar}[i]\in E_{aar}=\{Y0,Y1,N\}$, and (2) for all  $r_{ar}[|s|]\in E_{ar}=\{0,1,\varepsilon\}$, there exists $r_{a}[|s|]\in E_{A}=\{Y,N\}$ such that $r_{a}[|s|]r_{ar}[|s|]\in E_{aar}$ and $x_{v}=f_{v}(x_{v}',s[|s|]r_{a}[|s|]r_{ar}[|s|])=f_{v}(x_{0,v},s_{a}'s[|s|]r_{a}[|s|]r_{ar}[|s|])=f_{v}(x_{0,v},s[0]r_{a}[0]r_{ar}[0]s[1]$ $r_{a}[1]r_{ar}[1]\cdots s[|s|] r_{a}[|s|] r_{ar}[|s|])=f_{v}(x_{0,v},s_{a})$.
Since $G_{v}$ is part of $G_{aobs}$ and due to $|x_{v}^{3}|=1$,
by Definition~\ref{def:anonymity-string}, $r_{a}=r_{a}[0]r_{a}[1]\cdots r_{a}[|s|]$ is an anonymity-violating attack sequence for $s=s[0]s[1]\cdots s[|s|]$.
Thus, following Definition~\ref{def:anonymity-system}, $G$ violates current-state anonymity since there exists $s\in L(G)$ for which there exists an anonymity-violating sequence.}

{ (Only if) We illustrate that if the system violates current-state anonymity under $(X_{A},D)$, then $G_{v}$ is not an empty automaton.

By Definition~\ref{def:anonymity-system}, if the system violates current-state anonymity, there exists $s\in L(G)$ for which there exists an anonymity-violating sequence.
Suppose that the anonymity-violating sequence is denoted by $r_{a}=r_{a}[0]r_{a}[1]\cdots r_{a}[|s|]$. 
According to Definition~\ref{def:anonymity-string} and Lemma~\ref{lem:attack-observer}, there exists $r_{ar}'=r_{ar}[0]r_{ar}[1]r_{ar}[2]\cdots r_{ar}[|s|-1]$, for all $r_{ar}[|s|]\in E_{ar}=\{0,1,\varepsilon\}$ satisfying $r_{a}[i]r_{ar}[i]\in E_{aar}=\{Y0,Y1,N\}$ with $i=0,1,\cdots,|s|$ and $ 
   x_{aobs}=f_{aobs}(x_{0,aobs},s_{a})\neq \emptyset$ with $s_{a}=s[0]r_{a}[0]r_{ar}[0]s[1]r_{a}[1]r_{ar}[1]s[2]r_{a}[2]r_{ar}[2]\cdots s[|s|]$ $r_{a}[|s|]r_{ar}[|s|]$, we have $|x_{aobs}^{3}|=1$.
Therefore, for $s$ and $r_{a}$, we have $s_{a}\in L(G_{aobs})$, $x_{aobs}\in X_{aobs}$, and a sequence composed by states and events $x_{0,aobs}s[0]r_{a}[0]r_{ar}[0]x_{1,aobs}$ $s[1]r_{a}[1]r_{ar}[1]\cdots x_{|s|-1,aobs}s[|s|-1]r_{a}[|s|-1]r_{ar}[|s|-1]$ $x_{|s|-1,aobs}s[|s|]r_{a}[|s|]r_{ar}[|s|]x_{|s|,aobs}$ with $x_{|s|,aobs}=x_{aobs}$.
Thanks to $|x_{aobs}^{3}|=1$, $x_{aobs}$ is a type-I anonymity-violating state by Definition~\ref{def:anonymity-violation-state} and thus $x_{aobs}$ is a starting intermediate anonymity-violating type-I state. 
Then, by Definitions~\ref{def:intermediate-anonymity-violation-state-2}, \ref{def:intermediate-anonymity-violation-state-3}, and \ref{def:intermediate-anonymity-violation-state-1}, and Algorithm~\ref{alg:anonymity-violating-states}, we know that $x_{i,aobs}$ with $i\in \{1,2,\cdots, |s|\}$ is an intermediate anonymity-violating type-I state and $x_{0,aobs}$ is an intermediate anonymity-violating type-III state.
Finally, based on Definition~\ref{def:verifier}, $s_{a}\in L(G_{v})$ and $G_{v}$ is not an empty automaton. 
The proof is completed.
}
 $\hfill \blacksquare$
\end{thm}

\begin{myeg}\label{eg:composition-verifier-anonymity-violation}
(Example~\ref{eg:anonymity-violating-state} continued)
Following Definition~\ref{def:verifier}, the verifier is depicted in Fig.~\ref{fig:composition-verifer}.
By Theorem~\ref{thm:anonymity-violation}, the system violates current-state anonymity under a bounded number of state attacks with $X_{A}=\{2,4\}$ and $D=1$ since the verifier is not an empty automaton. 
 This conclusion aligns with Example~\ref{eg:current-state-anonymity-violating-string}, which argued that when the intruder performs a state attack after $a$ is observed, the intruder will definitely know the exact current state of the system.
    $\hfill \triangle$
\end{myeg}

\begin{figure}[!htbp] 
\centering
\scalebox{0.58}{ 

\begin{tikzpicture}[
  sstate/.style={rectangle, draw, minimum width=2cm, minimum height=0.5cm, fill=red!10, align=center},
  astate/.style={rectangle, draw, minimum width=2cm, minimum height=0.5cm, fill=green!10, align=center},
   aystate/.style={rectangle, draw, minimum width=2cm, minimum height=0.5cm, fill=blue!10, align=center},
  emptycell/.style={draw=none, fill=none, minimum width=2cm, minimum height=0.5cm},
  labelstyle/.style={midway, above, font=\small, fill=white, inner sep=1pt},
  initial text=,
  initial distance=1cm
]

\matrix (m) [matrix of nodes, row sep=1.2cm, column sep=0.8cm, nodes in empty cells] {
  \node[astate] (s6) {$(A,0,\{1,10\})$}; &
  \node[sstate] (s7) {$(S,0N,\{1,10\})$}; &
  \node[astate] (s8) {$(A,0,\{2,3\})$}; &
  \node[sstate] (s9) {$(S,0N,\{2,3\})$}; &\\
  \node[aystate] (s14) {$(AY,0Y,\{2,3\})$}; &
  \node[aystate] (s15) {$(AY,0Y,\{4,5\})$}; &
  \node[astate] (s16) {$(A,0,\{4,5\})$}; &
  \node[sstate] (s17) {$(S,0N,\{4,5\})$}; \\
  \node[sstate] (s20) {$(S,1,\{3\})$}; &
  \node[sstate] (s21) {$(S,1,\{2\})$}; &
  \node[sstate] (s22) {$(S,1,\{5\})$}; &
  \node[sstate] (s23) {$(S,1,\{4\})$}; \\
  \node[astate] (s26) {$(A,1,\{5\})$}; &
  \node[astate] (s27) {$(A,1,\{4\})$}; &
  \node[sstate] (s28) {$(S,1N,\{5\})$}; &
  \node[sstate] (s29) {$(S,1N,\{4\})$};  \\
};

\draw[->] (s14) -- node[labelstyle,left] {$0$} (s20);
\draw[->] (s15) -- node[labelstyle,left] {$0$} (s22);

\draw[->] (s14) -- node[labelstyle,left] {$1$} (s21);
\draw[->] (s15) -- node[labelstyle,left] {$1$} (s23);

\draw[->] (s8) -- node[labelstyle,left] {$Y$} (s14);

\draw[->] (s16) -- node[labelstyle,above] {$Y$} (s15);

\draw[->] (s6) -- node[labelstyle] {$N$} (s7);
\draw[->] (s8) -- node[labelstyle] {$N$} (s9);
\draw[->] (s16) -- node[labelstyle] {$N$} (s17);
\draw[->,bend right=20] (s26) to node[labelstyle] {$N$} (s28);
\draw[->,bend right=20] (s27) to node[labelstyle] {$N$} (s29);

\draw[->] (s7) -- node[labelstyle] {$a$} (s8);

\draw[->] (s9) -- node[labelstyle] {$b,c$} (s16);

\draw[->] (s20) to node[labelstyle] {$b$} (s26);
\draw[->] (s21) to node[labelstyle] {$b$} (s27);

\draw[->] (s20) to node[labelstyle,left,xshift=-0.5cm] {$c$} (s27);
\draw[->] (s21) to node[labelstyle,right,xshift=0.5cm] {$c$} (s26);
\draw[->] (s28) to node[labelstyle,left] {$c$} (s27);
\draw[->,bend left=25] (s29) to node[labelstyle,right] {$c$} (s26);
\draw[->,bend left] (s17) to node[labelstyle,right] {$c$} (s16);

\draw[->] (s22) to node[labelstyle,right,yshift=0.5cm] {$c$} (s27);
\draw[->] (s23.south) to node[labelstyle,below,xshift=2cm] {$c$} (s26.north);
\draw[->] ([xshift=-0.5cm]s6.west) -- (s6.west);

\end{tikzpicture}
}
\caption{\label{fig:composition-verifer}Verifier for the attack-observer in Fig. \ref{fig:composition} under  $X_{A}=\{2,4\}$ and $D=1$.}
\end{figure}
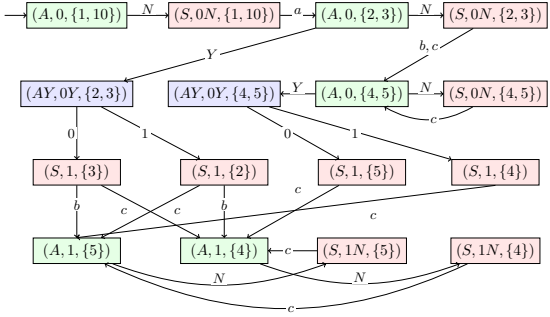

\section{Verification of attack-enforced anonymity violation}

{This section aims to verify whether an anonymity-violating system satisfies attack-enforced anonymity violation under a bounded number of state attacks, i.e., the existence of an attack strategy that guarantees that an anonymity-violating state will be reached regardless of the system activities.

Similar with the partition of states for the attack-observer, the states of the verifier can also be partitioned into three types of states: $X_{v}=X_{v1}~\dot{\cup}~X_{v2}~\dot{\cup}~X_{v3}$, where $X_{v1}= \{x_{v}\mid x_{v}^{1}=S \}$ is the set of type-I states, $X_{v2}= \{x_{v}\mid x_{v}^{1}=AY\}$ is the set of type-II states, and $X_{v3}= \{x_{v}\mid x_{v}^{1}=A\}$ is the set of type-III states.

\begin{mydef}
 \label{def:anonymity-vulnerable-state-uncontrollable} (Anonymity-vulnerable type-I state in an $(X_{A},D)$-verifier) Consider a fully-observable NFA  $G=(X,E,\delta,X_{0})$  under a bounded number of state attacks $(X_{A},D)$.
 Let $G_{v}=(X_{v},E_{v},f_{v},x_{0,v})$ be the corresponding $(X_{A},D)$-verifier.  
 A type-I state $x_{v1}=(x_{v1}^{1},x_{v1}^{2},x_{v1}^{3})\in X_{v1}$ is said to be an \emph{anonymity-vulnerable state} if
 \[(\forall e\in T_{aobs}(x_{v1}))[x_{v3}=f_{v}(x_{v1},e)\in X_{v3}],\]
 where $T_{aobs}$ denotes that we consider $G_{aobs}$ as the underlying automaton.
The set of anonymity-vulnerable type-I states is denoted by $X_{v1}^{avl}$.
    $\hfill \square$
\end{mydef}

\begin{mydef}
 \label{def:anonymity-vulnerable-state-uncontrollable-1} (Anonymity-vulnerable type-II state in an $(X_{A},D)$-verifier) Consider a fully-observable NFA  $G=(X,E,\delta,X_{0})$  under a bounded number of state attacks $(X_{A},D)$.
 Let $G_{v}=(X_{v},E_{v},f_{v},x_{0,v})$ be the corresponding $(X_{A},D)$-verifier.   
 A type-II state $x_{v2}=(x_{v2}^{1},x_{v2}^{2},x_{v2}^{3})\in X_{v2}$ is said to be an \emph{anonymity-vulnerable state} if
 \[(\forall \sigma\in T_{v}(x_{v2}))[x_{v1}=f_{v}(x_{v2},\sigma)\in X_{v1}^{avl}],\]
  where $T_{v}$ denotes that we consider $G_{v}$ as the underlying automaton.
The set of anonymity-vulnerable type-II states is denoted by $X_{v2}^{avl}$.
    $\hfill \square$
\end{mydef}

\begin{mydef}
 \label{def:anonymity-vulnerable-state-uncontrollable-2} (Anonymity-vulnerable type-III state in an $(X_{A},D)$-verifier) Consider a fully-observable NFA  $G=(X,E,\delta,X_{0})$  under a bounded number of state attacks $(X_{A},D)$.
 Let $G_{v}=(X_{v},E_{v},f_{v},x_{0,v})$ be the corresponding $(X_{A},D)$-verifier.   
 A type-III state $x_{v3}=(x_{v3}^{1},x_{v3}^{2},x_{v3}^{2})\in X_{v3}$ is said to be an \emph{anonymity-vulnerable state} if
 \[(\exists \sigma\in \{Y,N\})
  [x_{v}'=f_{v}(x_{v3},\sigma)\in  (X_{v1}^{avl}\cup X_{v2}^{avl})].\]
 The set of anonymity-vulnerable type-III states is denoted by $X_{v3}^{avl}$.
    $\hfill \square$
\end{mydef}

\begin{myeg}
    \label{eg:good-system-anonymity-violation-uncontrollable}
    (Example~\ref{eg:composition-verifier-anonymity-violation} continued)
    Reconsider the verifier in Fig.~\ref{fig:composition-verifer}.
   Given state $x_{v1}=\{(S,0N,\{1,10\})\}$, $x_{v1}$ is not anonymity-vulnerable by Definition~\ref{def:anonymity-vulnerable-state-uncontrollable} since for $d\in T_{aobs}(x_{v1})=$ $\{a,d\}$, we cannot find  a state $x_{v3}\in X_{v3}$ satisfying $x_{v3}=f_{v}(x_{v1},d)$.
 By Definition~\ref{def:anonymity-vulnerable-state-uncontrollable-1}, $\{(A,0,\{1,10\})\}$ is not anonymity-vulnerable. 
    $\hfill \triangle$
\end{myeg}

Next, a final verifier is constructed for collecting all anonymity-vulnerable states that are reachable from the initial state of the verifier, as the following algorithm illustrates.

\begin{algorithm}[!htbp]
    \caption{Construction of a final verifier.}\label{alg:final-verifier-uncontrollable}
    \LinesNumbered
    \KwIn{ A verifier $G_{v}=(X_{v},E_{v},f_{v},x_{0,v})$ for a current-state anonymity-violating NFA  $G=(X,E,\delta,X_{0})$ under a bounded number of state attacks $(X_{A},D)$. }
    \KwOut{ A final verifier $G_{fv}=(X_{fv},E_{fv},f_{fv},x_{0,fv})$.}
    Initialize $G_{fv}=G_{v}$\;
    Compute the set of anonymity-vulnerable type-I states $X_{v1}^{avl}$ by Definition~\ref{def:anonymity-vulnerable-state-uncontrollable}\;
    \While{There exists a type-I state that is not anonymity-vulnerable in $G_{v}$}{
    Obtain $G_{fv}=(X_{fv},E_{fv},f_{fv},x_{0,fv})$ by pruning all type-I states that are not anonymity-vulnerable and keeping the accessible part of $G_{v}$\;
    Let $G_{v}=G_{fv}$\;
      Compute the set of anonymity-vulnerable type-II states $X_{v2}^{avl}$ by Definition~\ref{def:anonymity-vulnerable-state-uncontrollable-1}\;
       Obtain $G_{fv}=(X_{fv},E_{fv},f_{fv},x_{0,fv})$ by pruning all type-II states that are not anonymity-vulnerable and keeping the accessible part of $G_{v}$\;
       Let $G_{v}=G_{fv}$\;
       Compute the set of anonymity-vulnerable type-III states $X_{v3}^{avl}$ by Definition~\ref{def:anonymity-vulnerable-state-uncontrollable-2}\;
     Obtain $G_{fv}=(X_{fv},E_{fv},f_{fv},x_{0,fv})$ by pruning type-III states that are not anonymity-vulnerable and keeping the accessible part of $G_{v}$\;
      Let $G_{v}=G_{fv}$\;}
     Return $G_{fv}=(X_{fv},E_{fv},f_{fv},x_{0,fv})$.
\end{algorithm}

\begin{mycom}
From Algorithm~\ref{alg:final-verifier-uncontrollable}, in the worst-case scenario, the numbers of states and edges in the obtained final verifier are identical with those in the verifier.
The anonymity-vulnerable states are computed iteratively, resulting in complexity of    $O((((2D+1)|E|+4D+2)2^{|X|})^{2})$.  
\end{mycom}

The set of states in a final verifier can also be partitioned to three types of states: $X_{fv}=X_{fv1}\dot{\cup} X_{fv2}\dot{\cup}X_{fv3}$, where $X_{fv1}=\{x_{fv}=(x_{fv}^{1},x_{fv}^{2},x_{fv}^{3})\mid x_{fv}^{1}=S\}$,  $X_{fv2}=\{x_{fv}=(x_{fv}^{1},x_{fv}^{2},x_{fv}^{3})\mid x_{fv}^{1}=AY\}$, and $X_{fv3}=\{x_{fv}=(x_{fv}^{1},x_{fv}^{2},x_{fv}^{3})\mid x_{fv}^{1}=A\}$.

\begin{thm}\label{thm:verification-system-attack-uncontrollable}
(Verification for attack-enforced anonymity violation)
Consider a fully-observable NFA  $G=(X,E,\delta, X_{0})$ under a bounded number of state attacks $(X_{A},D)$. 
Let $G_{fv}=(X_{fv},E_{fv},f_{fv},x_{0,fv})$ be the corresponding  final verifier.
System $G$ satisfies  attack-enforced anonymity violation if  and only if $G_{fv}$ is not an empty automaton.

{ Proof:  (If) We show that if $G_{fv}$ is not an empty automaton, the system $G$ satisfies attack-enforced anonymity violation under $(X_{A},D)$.

If $G_{fv}$ is not an empty automaton, by Algorithm~\ref{alg:final-verifier-uncontrollable}, for all $x_{fv}\in X_{fv}$, $x_{fv}$ is an anonymity-vulnerable state.
Since $G_{fv}$ is part of $G_{v}$, $x_{0,fv}\in X_{fv3}$ is an anonymity-vulnerable type-III state.
Then, based on Definition~\ref{def:anonymity-vulnerable-state-uncontrollable-2}, 
there exists $\sigma\in \{Y,N\}$ and there exists an anonymity-vulnerable state $x_{fv}'\in X_{fv1}\cup X_{fv2}$ such that $x_{fv}'=f_{fv}(x_{0,fv},\sigma)$.
If $x_{fv}'$ is an anonymity-vulnerable type-I state, i.e., $x_{fv}'\in X_{fv1}$, then we directly find the new anonymity-vulnerable type-I state.
If $x_{fv}'$ is an anonymity-vulnerable type-II state, i.e., $x_{fv}'\in X_{fv2}$, then by Definition~\ref{def:anonymity-vulnerable-state-uncontrollable-1},  
for all $\sigma\in T_{v}(x_{fv}')$, there exists an anonymity-vulnerable type-I state $x_{fv1}\in X_{fv1}$ such that $x_{fv1}=f_{fv}(x_{fv}',\sigma)$; thus we find the new anonymity-vulnerable type-I state.
For each case, we can find a new anonymity-vulnerable type-I state, denoted by $x_{fv1}$.
By Definition~\ref{def:anonymity-vulnerable-state-uncontrollable}, for the new anonymity-vulnerable  type-I state $x_{fv1}$, the following condition holds: for all $e\in T_{aobs}(x_{fv1}^{3})$, there exists an anonymity-vulnerable type-III state $x_{fv3}\in X_{fv3}$ such that $x_{fv3}=f_{fv}(x_{fv1},e)$.

Note that the number of anonymity-vulnerable states is bounded by $2^{|X|}(4D+2)$. By iteratively finding  new (different) anonymity-vulnerable  type-III states until no new anonymity-vulnerable  type-III state can be found (which also means that there is no new anonymity-vulnerable type-I or type-II state), for all $s\in L(G)$, the attack sequence $r_{a}$ for $s$ is listed in $G_{fv}$.
Since $G_{fv}$ is part of $G_{v}$, and $G_{fv}$ and $G_{v}$ are both DFAs, for any $s_{a}\in L(G_{fv})$ obtained from $s$ and $r_{a}$ in $G_{fv}$, we know that $s_{a}\in L(G_{v})$ and the visited sequence of states are identical.
Since $G_{v}$ collects all $s\in L(G)$ such that  $r_{a}$ is an anonymity-violating attack sequence for $s$, we can let $s=s'es''$ and $r_{a}=r_{a}'r_{a}'''r_{a}''$ such that $r_{a}'$, $r_{a}'''$, and $r_{a}''$ are the attack sequence for $s'$, $e$, and $s''$, respectively.
Following Definition~\ref{def:anonymity-attackable-string}, $r_{a}'$ is an attack-enforced anonymity-violating attack sequence for $s'$.
When splitting $s$ into $s'es''$, the length of $|s''|$ can be from 0 to $|s|$ (resp. the length of $|s'|$ can be from $|s|$ to 0), and correspondingly,  $r_{a}'$, $r_{a}'''$, and $r_{a}''$ can be adjusted to be consistent with $s'$, $e$, and $s''$ such that $r_{a}=r_{a}'r_{a}'''r_{a}''$.
Therefore, based on Definition~\ref{def:anonymity-attackable}, the system $G$ is attack-enforced anonymity-violating under $(X_{A},D)$.}

{ (Only if) We illustrate that $G$ satisfies attack-enforced anonymity violation under $(X_{A},D)$ if $G_{fv}$ is not an empty automaton. By contradiction, if $G_{fv}$ is an empty automaton, $G$ does not satisfy attack-enforced anonymity violation.

Since $G_{fv}$ is an empty automaton and $G_{fv}$ is obtained from $G_{v}$ by Algorithm~\ref{alg:final-verifier-uncontrollable}, it has the following two scenarios: (1) $G_{v}$ is an empty automaton, and (2) $G_{v}$ is not an empty automaton.
Note that $G_{v}$ collects all $s\in L(G)$ such that there exists an anonymity-violating attack sequence $r_{a}$ for $s$.
For scenario (1), there does not exist $s\in L(G)$ and $r_{a}$ such that $r_{a}$ is an anonymity-violating attack sequence for $s$. 
By Definition~\ref{def:anonymity-attackable-string}, there does not exist $s$ and $r_{a}$ such that $r_{a}$ is an attack-enforced anonymity-violating sequence.
Therefore, as described in Definition~\ref{def:anonymity-attackable}, $G$ does not satisfy attack-enforced anonymity violation.

For scenario (2), there necessarily exist $s\in L(G)$ and $r_{a}$ such that $r_{a}$ is an anonymity-violating attack sequence for $s$.
Following Algorithm~\ref{alg:final-verifier-uncontrollable}, there necessarily exists a type-I state that is not anonymity-vulnerable  since $G_{fv}$ is an empty automaton and $G_{v}$ is not an empty automaton.
Let's consider an arbitrary type-I state that is not anonymity-vulnerable, denoted by $x_{v1}\in X_{v1}$.
Based on Definition~\ref{def:anonymity-vulnerable-state-uncontrollable}, there exists $e\in T_{aobs}(x_{v1})$ such that we cannot find a type-III state $x_{v3}\in X_{v3}$ satisfying $x_{v3}=f_{v}(x_{v1},e)$.
Therefore, there exists $s\in L(G)$ satisfying $x_{v1}^{3}=f_{v}(x_{0,v}^{3},s)$,
and for $s$, there exists $e\in E$ such that $r_{a}''$ cannot be found for $e$.
For any $s_{a}\in L(G_{v})$ such that $x_{v3}=f_{v}(x_{0,v},s_{a})$, for $s$, we can obtain $r_{a}$ from $s_{a}$.
By Definition~\ref{def:anonymity-attackable-string}, 
$r_{a}$ is not an attack-enforced anonymity-violating attack sequence for $s$.
Following Definition~\ref{def:anonymity-attackable}, $G$ is not attack-enforced anonymity-violating since there exists $s\in L(G)$ such that there does not exist an attack-enforced anonymity-violating attack sequence.

For all scenarios, $G$ is not attack-enforced anonymity-violating, resulting that $G$ is not attack-enforced anonymity-violating. 
Finally, the proof is completed.
}
    $\hfill \blacksquare$
\end{thm}

\begin{myeg}
    \label{final-verifier}
    (Example~\ref{eg:good-system-anonymity-violation-uncontrollable} continued)
    Following Algorithm~\ref{alg:final-verifier-uncontrollable}, the obtained final verifier is the empty automaton.
Therefore, the considered NFA in Fig.~\ref{fig:system} under $X_{A}=\{2,4\}$ and $D=1$ does not satisfy attack-enforced anonymity violation.
    $\hfill \triangle$
\end{myeg}

\begin{myeg}
\label{eg:attack-enforced-running-example}
Let us consider the NFA in Fig.~\ref{fig:system}. Suppose that $X_{A}=\{2,4,8,9\}$ and $D=1$.
According to Algorithm~\ref{alg:verification-algorithm}, the observation model and the verifier are obtained as shown in Figs.~\ref{fig:composition-2469} and \ref{fig:verifier-2469}, respectively, and the NFA under a bounded number of state attacks with $X_{A}=\{2,4,8,9\}$ and $D=1$ satisfies current-state anonymity violation.
According to Algorithm~\ref{alg:final-verifier-uncontrollable}, the final verifier $G_{fv}$ is obtained in Fig.~\ref{fig:final-verifier-2469}.
By Theorem~\ref{thm:verification-system-attack-uncontrollable}, the system satisfies attack-enforced anonymity violation.
    $\hfill \triangle$
\end{myeg}

}

\begin{figure*}[!htbp] 
\centering
\scalebox{0.56}{

\begin{tikzpicture}[
  sstate/.style={rectangle, draw, minimum width=2cm, minimum height=0.5cm, fill=red!10, align=center},
  astate/.style={rectangle, draw, minimum width=2cm, minimum height=0.5cm, fill=green!10, align=center},
   aystate/.style={rectangle, draw, minimum width=2cm, minimum height=0.5cm, fill=blue!10, align=center},
  emptycell/.style={draw=none, fill=none, minimum width=2cm, minimum height=0.5cm},
  labelstyle/.style={midway, above, font=\small, fill=white, inner sep=1pt},
  initial text=,
  initial distance=1cm
]

\matrix (m) [matrix of nodes, row sep=1.2cm, column sep=0.8cm, nodes in empty cells] {
 \node[sstate] (s0y) {$(S,1,\{9\})$}; &
  \node[astate] (s1y) {$(A,1,\{7\})$}; &
  \node[sstate] (s2y) {$(S,1N,\{7\})$}; &
  \node[astate] (s3y) {$(A,1,\{8\})$}; &
  \node[sstate] (s4y) {$(S,1N,\{8\})$}; &
  \node[sstate] (s5y) {$(S,1,\{8\})$}; \\
  \node[sstate] (s0) {$(S,1,\{6\})$}; &
  \node[aystate] (s1) {$(AY,0Y,\{6,9\})$}; &
  \node[astate] (s2) {$(A,0,\{6,9\})$}; &
  \node[sstate] (s3) {$(S,0N,\{6,9\})$}; &
  \node[astate] (s4) {$(A,0,\{7,8\})$}; &
  \node[sstate] (s5) {$(S,0N,\{7,8\})$}; \\
  \node[astate] (s6) {$(A,0,\{1,10\})$}; &
  \node[sstate] (s7) {$(S,0N,\{1,10\})$}; &
  \node[astate] (s8) {$(A,0,\{2,3\})$}; &
  \node[sstate] (s9) {$(S,0N,\{2,3\})$}; &
  \node[aystate] (s10) {$(AY,0Y,\{7,8\})$}; &
  \node[sstate] (s11) {$(S,1,\{7\})$}; \\
  \node[aystate] (s12) {$(AY,0Y,\{1,10\})$}; &
  \node[sstate] (s13) {$(S,1N,\{4,5\})$}; &
  \node[aystate] (s14) {$(AY,0Y,\{2,3\})$}; &
  \node[aystate] (s15) {$(AY,0Y,\{4,5\})$}; &
  \node[astate] (s16) {$(A,0,\{4,5\})$}; &
  \node[sstate] (s17) {$(S,0N,\{4,5\})$}; \\
  \node[sstate] (s18) {$(S,1,\{1,10\})$}; &
  \node[astate] (s19) {$(A,1,\{4,5\})$}; &
  \node[sstate] (s20) {$(S,1,\{3\})$}; &
  \node[sstate] (s21) {$(S,1,\{2\})$}; &
  \node[sstate] (s22) {$(S,1,\{5\})$}; &
  \node[sstate] (s23) {$(S,1,\{4\})$}; \\
  \node[astate] (s24) {$(A,1,\{2,3\})$}; &
  \node[sstate] (s25) {$(S,1N,\{2,3\})$}; &
  \node[astate] (s26) {$(A,1,\{5\})$}; &
  \node[astate] (s27) {$(A,1,\{4\})$}; &
  \node[sstate] (s28) {$(S,1N,\{5\})$}; &
  \node[sstate] (s29) {$(S,1N,\{4\})$}; \\
  \node[astate] (s30) {$(A,1,\{6,9\})$}; &
  \node[sstate] (s31) {$(S,1N,\{6,9\})$}; &
  \node[astate] (s32) {$(A,1,\{7,8\})$}; &
  \node[sstate] (s33) {$(S,1N,\{7,8\})$}; &
  \node[emptycell] {}; &
  \node[emptycell] {}; \\
};

\draw[->] (s1) -- node[labelstyle] {$0$} (s0);
\draw[->] (s10) -- node[labelstyle] {$0$} (s11);
\draw[->] (s12) -- node[labelstyle,left] {$0$} (s18);
\draw[->] (s14) -- node[labelstyle,left] {$0$} (s20);
\draw[->] (s15) -- node[labelstyle,left] {$0$} (s22);

\draw[->] (s14) -- node[labelstyle,left] {$1$} (s21);
\draw[->] (s15) -- node[labelstyle,left] {$1$} (s23);
\draw[->] (s1) -- node[labelstyle,left] {$1$} (s0y);
\draw[->] (s10) -- ++ (5cm,1.5cm) -- node[labelstyle,left] {$1$} (s5y.east);

\draw[->] (s2) -- node[labelstyle] {$Y$} (s1);
\draw[->] (s6) -- node[labelstyle,left] {$Y$} (s12);
\draw[->] (s8) -- node[labelstyle,left] {$Y$} (s14);
\draw[->] (s4) -- node[labelstyle,left] {$Y$} (s10);
\draw[->] (s16) -- node[labelstyle,above] {$Y$} (s15);

\draw[->] (s2) -- node[labelstyle] {$N$} (s3);
\draw[->] (s4) -- node[labelstyle] {$N$} (s5);
\draw[->] (s6) -- node[labelstyle] {$N$} (s7);
\draw[->] (s8) -- node[labelstyle] {$N$} (s9);
\draw[->] (s16) -- node[labelstyle] {$N$} (s17);
\draw[->] (s19) -- node[labelstyle] {$N$} (s13);
\draw[->] (s24) -- node[labelstyle] {$N$} (s25);
\draw[->,bend right=20] (s26) to node[labelstyle] {$N$} (s28);
\draw[->,bend right=20] (s27) to node[labelstyle] {$N$} (s29);
\draw[->] (s30) -- node[labelstyle] {$N$} (s31);
\draw[->] (s32) -- node[labelstyle] {$N$} (s33);
\draw[->] (s1y) -- node[labelstyle] {$N$} (s2y);
\draw[->] (s3y) -- node[labelstyle] {$N$} (s4y);

\draw[->] (s18) -- node[labelstyle,left] {$a$} (s24);
\draw[->] (s7) -- node[labelstyle] {$a$} (s8);

\draw[->,bend right] (s18.west) to node[labelstyle,right] {$d$} (s30.west);
\draw[->] (s7) -- node[labelstyle] {$d$} (s2);

\draw[->] (s25) -- node[labelstyle] {$b,c$} (s19);
\draw[->] (s9) -- node[labelstyle] {$b,c$} (s16);

\draw[->] (s31) to node[labelstyle] {$b$} (s32);
\draw[->]
  (s0.west)
  -- ++(-1cm, 0) -- ++(0,-9.8cm) -- ++ (8.89cm,0) 
  node[labelstyle] {$b$}
  -- (s32.south);
\draw[->] (s3) to node[labelstyle] {$b$} (s4);
\draw[->] (s20) to node[labelstyle] {$b$} (s26);
\draw[->] (s21) to node[labelstyle] {$b$} (s27);
\draw[->] (s0y) -- node[labelstyle] {$b$} (s1y);

\draw[->,bend left] (s33) to node[labelstyle] {$c$} (s32);
\draw[->,bend left] (s13) to node[labelstyle] {$c$} (s19);
\draw[->] (s20) to node[labelstyle,left,xshift=-0.5cm] {$c$} (s27);
\draw[->] (s21) to node[labelstyle,right,xshift=0.5cm] {$c$} (s26);
\draw[->] (s28) to node[labelstyle,left] {$c$} (s27);
\draw[->,bend left=25] (s29) to node[labelstyle,right] {$c$} (s26);
\draw[->,bend left] (s17) to node[labelstyle,right] {$c$} (s16);
\draw[->,bend left] (s5) to node[labelstyle,above] {$c$} (s4);
\draw[->] (s22) to node[labelstyle,right,yshift=0.5cm] {$c$} (s27);
\draw[->] (s23.south) to node[labelstyle,below,xshift=2cm] {$c$} (s26.north);
\draw[->] ([yshift=0.5cm]s6.north) -- (s6.north);
\draw[->] (s2y) to node[labelstyle,above] {$c$} (s3y);
\draw[->,bend right=10] (s4y) to node[labelstyle,above] {$c$} (s1y);
\draw[->,bend right=15] (s5y) to node[labelstyle,above] {$c$} (s1y);
\draw[->,bend left] (s11.east) -- ++  (1cm,0) -- ++ (0,3cm) -- ++ (-7cm,0) -- node[labelstyle,above] {$c$} (s3y);

\end{tikzpicture}
}
\caption{\label{fig:composition-2469}Observation model for the NFA in Fig. \ref{fig:system} under a bounded number of state attacks with $X_{A}=\{2,4,8,9\}$ and $D=1$.}
\end{figure*}
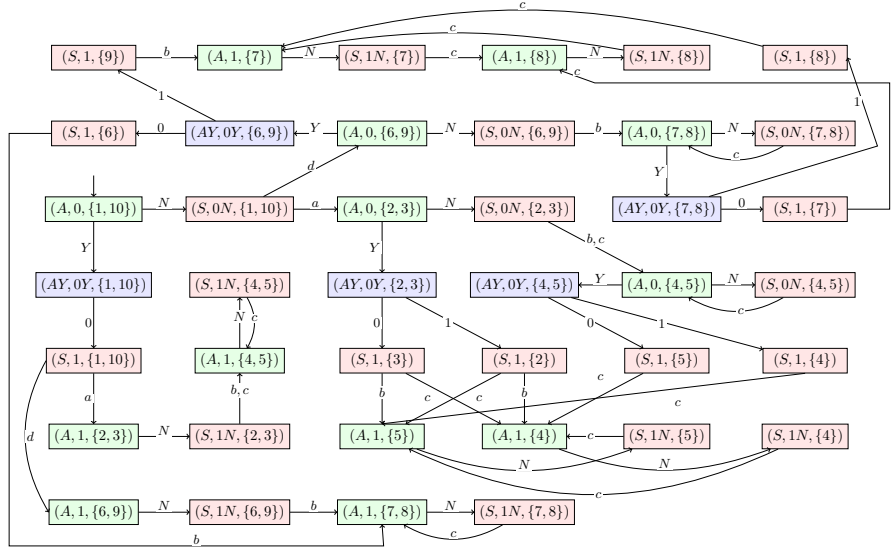

\begin{figure*}[!htbp] 
\centering
\scalebox{0.6}{

\begin{tikzpicture}[
  sstate/.style={rectangle, draw, minimum width=2cm, minimum height=0.5cm, fill=red!10, align=center},
  astate/.style={rectangle, draw, minimum width=2cm, minimum height=0.5cm, fill=green!10, align=center},
   aystate/.style={rectangle, draw, minimum width=2cm, minimum height=0.5cm, fill=blue!10, align=center},
  emptycell/.style={draw=none, fill=none, minimum width=2cm, minimum height=0.5cm},
  labelstyle/.style={midway, above, font=\small, fill=white, inner sep=1pt},
  initial text=,
  initial distance=1cm
]

\matrix (m) [matrix of nodes, row sep=1.3cm, column sep=0.8cm, nodes in empty cells] {
 \node[sstate] (s0y) {$(S,1,\{9\})$}; &
  \node[astate] (s1y) {$(A,1,\{7\})$}; &
  \node[sstate] (s2y) {$(S,1N,\{7\})$}; &
  \node[astate] (s3y) {$(A,1,\{8\})$}; &
  \node[sstate] (s4y) {$(S,1N,\{8\})$}; &
  \node[sstate] (s5y) {$(S,1,\{8\})$}; \\
  \node[sstate] (s0) {$(S,1,\{6\})$}; &
  \node[aystate] (s1) {$(AY,0Y,\{6,9\})$}; &
  \node[astate] (s2) {$(A,0,\{6,9\})$}; &
  \node[sstate] (s3) {$(S,0N,\{6,9\})$}; &
  \node[astate] (s4) {$(A,0,\{7,8\})$}; &
  \node[sstate] (s5) {$(S,0N,\{7,8\})$}; \\
  \node[astate] (s6) {$(A,0,\{1,10\})$}; &
  \node[sstate] (s7) {$(S,0N,\{1,10\})$}; &
  \node[astate] (s8) {$(A,0,\{2,3\})$}; &
  \node[sstate] (s9) {$(S,0N,\{2,3\})$}; &
  \node[aystate] (s10) {$(AY,0Y,\{7,8\})$}; &
  \node[sstate] (s11) {$(S,1,\{7\})$}; \\
 \node[sstate] (s20) {$(S,1,\{3\})$}; &
  \node[aystate] (s14) {$(AY,0Y,\{2,3\})$}; &
  \node[sstate] (s21) {$(S,1,\{2\})$}; &
  \node[aystate] (s15) {$(AY,0Y,\{4,5\})$}; &
  \node[astate] (s16) {$(A,0,\{4,5\})$}; &
  \node[sstate] (s17) {$(S,0N,\{4,5\})$}; \\
  \node[sstate] (s22) {$(S,1,\{5\})$}; &
  \node[sstate] (s23) {$(S,1,\{4\})$}; &
  \node[astate] (s26) {$(A,1,\{5\})$}; &
  \node[astate] (s27) {$(A,1,\{4\})$}; &
  \node[sstate] (s28) {$(S,1N,\{5\})$}; &
  \node[sstate] (s29) {$(S,1N,\{4\})$}; \\
};

\draw[->] (s1) -- node[labelstyle] {$0$} (s0);
\draw[->] (s10) -- node[labelstyle] {$0$} (s11);
\draw[->,bend right=20] (s14) -- node[labelstyle,left] {$0$} (s20);
\draw[->] (s15) -- node[labelstyle,below] {$0$} (s22);

\draw[->] (s14) -- node[labelstyle,left] {$1$} (s21);
\draw[->] (s15) -- node[labelstyle,right] {$1$} (s23);
\draw[->] (s1) -- node[labelstyle,left] {$1$} (s0y);
\draw[->] (s10) -- ++ (5cm,1.2cm) -- node[labelstyle,left] {$1$} (s5y.east);

\draw[->] (s2) -- node[labelstyle] {$Y$} (s1);
\draw[->] (s8) -- node[labelstyle,left] {$Y$} (s14);
\draw[->] (s4) -- node[labelstyle,left] {$Y$} (s10);
\draw[->] (s16) -- node[labelstyle,above] {$Y$} (s15);

\draw[->] (s2) -- node[labelstyle] {$N$} (s3);
\draw[->] (s4) -- node[labelstyle] {$N$} (s5);
\draw[->] (s6) -- node[labelstyle] {$N$} (s7);
\draw[->] (s8) -- node[labelstyle] {$N$} (s9);
\draw[->] (s16) -- node[labelstyle] {$N$} (s17);

\draw[->,bend right=20] (s26) to node[labelstyle] {$N$} (s28);
\draw[->,bend right=20] (s27) to node[labelstyle] {$N$} (s29);
\draw[->] (s1y) -- node[labelstyle] {$N$} (s2y);
\draw[->] (s3y) -- node[labelstyle] {$N$} (s4y);

\draw[->] (s7) -- node[labelstyle] {$a$} (s8);

\draw[->] (s7) -- node[labelstyle] {$d$} (s2);

\draw[->] (s9) -- node[labelstyle] {$b,c$} (s16);

\draw[->] (s3) to node[labelstyle] {$b$} (s4);
\draw[->] (s20) to node[labelstyle] {$b$} (s26);
\draw[->] (s21) to node[labelstyle] {$b$} (s27);
\draw[->] (s0y) -- node[labelstyle] {$b$} (s1y);

\draw[->] (s20) to node[labelstyle,above] {$c$} (s27);
\draw[->] (s21) to node[labelstyle,left] {$c$} (s26);
\draw[->] (s28) to node[labelstyle,left] {$c$} (s27);
\draw[->,bend left=25] (s29) to node[labelstyle,right] {$c$} (s26);
\draw[->,bend left] (s17) to node[labelstyle,right] {$c$} (s16);
\draw[->,bend left] (s5) to node[labelstyle,above] {$c$} (s4);
\draw[->,bend right=20] (s22) to node[labelstyle,right] {$c$} (s27);
\draw[->] (s23) to node[labelstyle,below] {$c$} (s26);
\draw[->] ([yshift=0.5cm]s6.north) -- (s6.north);
\draw[->] (s2y) to node[labelstyle,above] {$c$} (s3y);
\draw[->,bend right=15] (s4y) to node[labelstyle,above] {$c$} (s1y);
\draw[->,bend right=20] (s5y) to node[labelstyle,above] {$c$} (s1y);
\draw[->,bend left] (s11.east) -- ++  (0.8cm,0) -- ++ (0,3cm) -- ++ (-7cm,0) -- node[labelstyle,above] {$c$} (s3y);

\end{tikzpicture}
}
\caption{\label{fig:verifier-2469}Verifier of the attack-observer in  Fig. \ref{fig:composition-2469} under a bounded number of state attacks with $X_{A}=\{2,4,8,9\}$ and $D=1$.}
\end{figure*}
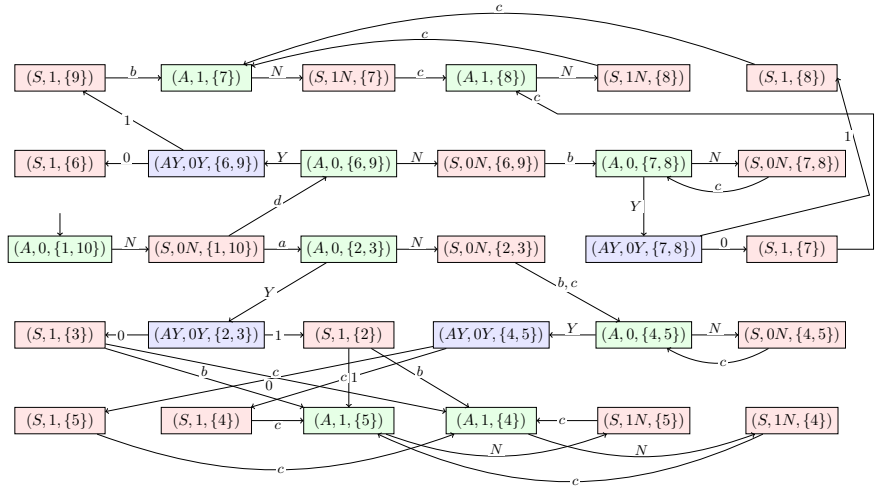

\begin{figure*}[!htbp] 
\centering
\scalebox{0.6}{

\begin{tikzpicture}[
  sstate/.style={rectangle, draw, minimum width=2cm, minimum height=0.5cm, fill=red!10, align=center},
  astate/.style={rectangle, draw, minimum width=2cm, minimum height=0.5cm, fill=green!10, align=center},
   aystate/.style={rectangle, draw, minimum width=2cm, minimum height=0.5cm, fill=blue!10, align=center},
  emptycell/.style={draw=none, fill=none, minimum width=2cm, minimum height=0.5cm},
  labelstyle/.style={midway, above, font=\small, fill=white, inner sep=1pt},
  initial text=,
  initial distance=1cm
]

\matrix (m) [matrix of nodes, row sep=1.2cm, column sep=0.8cm, nodes in empty cells] {
  \node[emptycell] {}; &
  \node[astate] (s1y) {$(A,1,\{7\})$}; &
  \node[sstate] (s2y) {$(S,1N,\{7\})$}; &
  \node[astate] (s3y) {$(A,1,\{8\})$}; &
  \node[sstate] (s4y) {$(S,1N,\{8\})$}; &
  \node[sstate] (s5y) {$(S,1,\{8\})$}; \\
  \node[astate] (s2) {$(A,0,\{6,9\})$}; &
  \node[sstate] (s3) {$(S,0N,\{6,9\})$}; &
  \node[astate] (s4) {$(A,0,\{7,8\})$}; &
  \node[sstate] (s5) {$(S,0N,\{7,8\})$}; 
   \node[emptycell] {}; &
   \node[emptycell] {}; &\\
  \node[astate] (s6) {$(A,0,\{1,10\})$}; &
  \node[sstate] (s7) {$(S,0N,\{1,10\})$}; &
  \node[astate] (s8) {$(A,0,\{2,3\})$}; &
  \node[sstate] (s9) {$(S,0N,\{2,3\})$}; &
  \node[aystate] (s10) {$(AY,0Y,\{7,8\})$}; &
  \node[sstate] (s11) {$(S,1,\{7\})$}; \\
 \node[sstate] (s20) {$(S,1,\{3\})$}; &
  \node[aystate] (s14) {$(AY,0Y,\{2,3\})$}; &
  \node[sstate] (s21) {$(S,1,\{2\})$}; &
  \node[aystate] (s15) {$(AY,0Y,\{4,5\})$}; &
  \node[astate] (s16) {$(A,0,\{4,5\})$}; &
  \node[sstate] (s17) {$(S,0N,\{4,5\})$}; \\
  \node[sstate] (s22) {$(S,1,\{5\})$}; &
  \node[sstate] (s23) {$(S,1,\{4\})$}; &
  \node[astate] (s26) {$(A,1,\{5\})$}; &
  \node[astate] (s27) {$(A,1,\{4\})$}; &
  \node[sstate] (s28) {$(S,1N,\{5\})$}; &
  \node[sstate] (s29) {$(S,1N,\{4\})$}; \\
};


\draw[->] (s10) -- node[labelstyle] {$0$} (s11);
\draw[->,bend right=20] (s14) -- node[labelstyle,left] {$0$} (s20);
\draw[->] (s15) -- node[labelstyle,below] {$0$} (s22);

\draw[->] (s14) -- node[labelstyle,left] {$1$} (s21);
\draw[->] (s15) -- node[labelstyle,right] {$1$} (s23);

\draw[->] (s10) -- node[labelstyle,left] {$1$} (s5y);

\draw[->] (s8) -- node[labelstyle,left] {$Y$} (s14);
\draw[->] (s4) -- node[labelstyle,left] {$Y$} (s10);
\draw[->] (s16) -- node[labelstyle,above] {$Y$} (s15);

\draw[->] (s2) -- node[labelstyle] {$N$} (s3);
\draw[->] (s4) -- node[labelstyle] {$N$} (s5);
\draw[->] (s6) -- node[labelstyle] {$N$} (s7);
\draw[->] (s8) -- node[labelstyle] {$N$} (s9);
\draw[->] (s16) -- node[labelstyle] {$N$} (s17);

\draw[->,bend right=20] (s26) to node[labelstyle] {$N$} (s28);
\draw[->,bend right=20] (s27) to node[labelstyle] {$N$} (s29);
\draw[->] (s1y) -- node[labelstyle] {$N$} (s2y);
\draw[->] (s3y) -- node[labelstyle] {$N$} (s4y);

\draw[->] (s7) -- node[labelstyle] {$a$} (s8);

\draw[->] (s7) -- node[labelstyle] {$d$} (s2);

\draw[->] (s9) -- node[labelstyle] {$b,c$} (s16);

\draw[->] (s3) to node[labelstyle] {$b$} (s4);
\draw[->] (s20) to node[labelstyle] {$b$} (s26);
\draw[->] (s21) to node[labelstyle] {$b$} (s27);

\draw[->] (s20) to node[labelstyle,above] {$c$} (s27);
\draw[->] (s21) to node[labelstyle,left] {$c$} (s26);
\draw[->] (s28) to node[labelstyle,left] {$c$} (s27);
\draw[->,bend left=25] (s29) to node[labelstyle,right] {$c$} (s26);
\draw[->,bend left] (s17) to node[labelstyle,right] {$c$} (s16);
\draw[->,bend right] (s5) to node[labelstyle,above] {$c$} (s4);
\draw[->,bend right=20] (s22) to node[labelstyle,right] {$c$} (s27);
\draw[->] (s23) to node[labelstyle,below] {$c$} (s26);
\draw[->] ([yshift=0.5cm]s6.north) -- (s6.north);
\draw[->] (s2y) to node[labelstyle,above] {$c$} (s3y);
\draw[->,bend right=10] (s4y) to node[labelstyle,above] {$c$} (s1y);
\draw[->,bend right=15] (s5y) to node[labelstyle,above] {$c$} (s1y);
\draw[->,bend left] (s11) -- node[labelstyle,above] {$c$} (s3y);

\end{tikzpicture}
}
\caption{\label{fig:final-verifier-2469} Final verifier for the NFA in Fig.~\ref{fig:system} under $(X_{A},D)=(\{2,4,8,9\},1)$.}
\end{figure*}
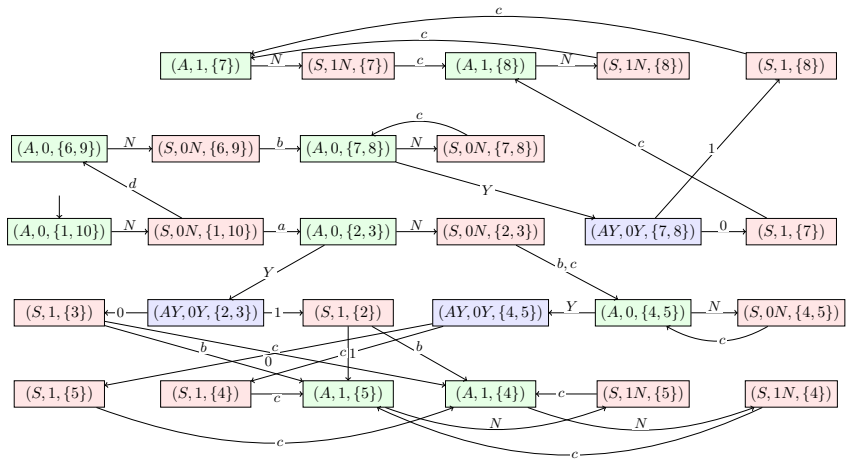

\section{Synthesis of an attack-enforced anonymity violation strategy}

If a fully-observable NFA satisfies attack-enforced anonymity violation under a bounded number of state attacks with respect to $X_{A}$ and $D$, based on the final verifier, we can  synthesize an attack strategy that guarantees that current-state anonymity will be violated as the system evolves.

\begin{mydef}
    \label{def:attack-enforced-game-structure}
(Attack-enforced anonymity violation strategy)    
Consider a fully-observable NFA $G=(X,E,\delta,X_{0})$, a set of attacked states $X_{A}$, and a maximum $D$ state attacks with respect to $X_{A}$.
Let $G_{fv}=(X_{fv},E_{fv},f_{fv},x_{0,fv})$  be the corresponding final verifier.
The attack-enforced anonymity violation strategy is  denoted by $G_{as}=(X_{as},E_{as},f_{as},x_{0,as})$,
where
\begin{enumerate}
    \item $X_{as}\subseteq \{x_{0,fv}\}\cup X_{fv1}$ is the set of states;
    \item $E_{as}=E^{+}/E_{arr} $ is the set of events, where $E^{+}=E\cup \{\varepsilon\}$ and $E_{arr} =\{Y0,Y1,N\}$;
    \item $x_{0,as}=x_{0,fv}$ is the initial state;
    \item $f_{as}$ is the transition function, defined as:\\ for all $x_{as}\in X_{as}$ and for all $e_{as}=e/\sigma\in E_{as}$, $f_{as}(x_{as},e/\sigma)=f_{fv}(x_{as},e\sigma)$.

\end{enumerate}
$\hfill \square$
\end{mydef}

An  attack-enforced anonymity violation strategy, denoted by $G_{as}=(X_{as}, E_{as}, f_{as}, x_{0,as})$,  is  derived from the final verifier and is represented by a Mealy automaton\cite{Mealy-automaton},
which can be obtained by Algorithms~\ref{alg:call}  and~\ref{alg:systhesis-attack-strategy-algorithm}  below.

\begin{algorithm}[!htbp]
    \caption{$Func(x_{as}, e, e_{a})$.}\label{alg:call}
    \LinesNumbered
    \KwIn{Current examined state $x_{as}$, current examined event $e$,  and current examined attack action $e_{a}$. }
    \KwOut{None.}
  $x_{fv}=f_{fv}(x_{as},e)$\;
 \If{$e_{a}=N$}{
  $x_{as}'=f_{fv}(x_{fv},e_{a})$\;
   Add $f_{as}(x_{as},e/ e_{a})=x_{as}'$ to $f_{as}$\;
   \If{$x_{as}'\notin X_{as}$}{
  Add $x_{as}'$ to $X_{as}$\;
 } }
 \Else{
 $x_{fv}'=f_{fv}(x_{fv},e_{a})$\;
 \For{$\sigma
  \in T_{fv}(x_{fv}')$ that has not been examined}{
  $x_{as}'=f_{fv}(x_{as},ee_{a}\sigma)$\;
 Add $f_{as}(x_{as},e/ e_{a} \sigma)=x_{as}'$ to $f_{as}$\;
   \If{$x_{as}'\notin X_{as}$}{
  Add $x_{as}'$ to $X_{as}$\;
  }
  Mark $\sigma$ is examined.
  }
 }
\end{algorithm}

\begin{myeg}
\label{eg:attack-enforced-anonymity-violation-strategy-func}
(Example~\ref{eg:attack-enforced-running-example} continued)
Consider the final verifier in Fig.~\ref{fig:final-verifier-2469}.
Suppose that $x_{as}=(S,0N,\{2,3\})$, $e=b$, and $e_{a}=Y$.
We  describe how Algorithm~\ref{alg:call} works. 
In Line~1, we have $x_{fv}=(A,0,\{4,5\})$.
Since $e_{a}=Y$, we go to lines~8--14.
Then, $x_{fv}'=(AY,0Y,\{4,5\})$.
For $\sigma=0,1\in T_{fv}(x_{fv}')=\{0,1\}$, we have $x_{as}'=f_{fv}((S,0N,\{2,3\}),bY0)=(S,1,\{5\})$ (resp. $x_{as}'=f_{fv}((S,0N,\{2,3\}),bY1)=(S,1,\{4\})$) and add $x_{as}'=f_{as}((S,0N,\{2,3\}),b/Y0)=(S,1,\{5\})$ (resp. $x_{as}'=f_{as}((S,0N,\{2,3\}),b/Y1)=(S,1,\{4\})$) to $f_{as}$.

If $e_{a}=N$, then we go to lines~2--6.
Subsequently, we have  $x_{as}'=(S,0N,\{4,5\})$ and add $x_{as}'=f_{as}((S,0N,\{2,3\})$, $b/N)=(S,0N,\{4,5\})$ to $f_{as}$.
    $\hfill \triangle$
\end{myeg}

\begin{algorithm}[!htbp]
    \caption{Construction of an attack-enforced anonymity violation strategy.}\label{alg:systhesis-attack-strategy-algorithm}
    \LinesNumbered
    \KwIn{ A final verifier $G_{fv}=(X_{fv},E_{fv},f_{fv},x_{0,fv})$  for an  NFA  $G=(X,E,\delta,X_{0})$ satisfying attack-enforced anonymity violation under a bounded number of state attacks $(X_{A},D)$. }
    \KwOut{An attack-enforced anonymity violation strategy $G_{as}=(X_{as}, E_{as}, f_{as}, x_{0,as})$.}
     Initialize $x_{0,as}=x_{0,fv}$ and  $X_{as}=\{x_{0,as}\}$\;
      Choose $e_{a}\in \{Y,N\}$ satisfying $e_{a}\in T_{fv}(x_{0,as})$\;
       
Let $e_{a}$ be the choice for $x_{0,as}$ and let $e=\varepsilon$\;
$Func(x_{0,as}, e, e_{a})$\;
  Mark that $x_{0,as}$ is examined\;
  \For{$x_{as}\in X_{as}$ that has not been examined}{
  \For{$e\in T_{aobs}(x_{as}^{3})\cap E$ that has not been examined}{
  Choose $e_{a}\in \{Y,N\}$ to be the attack action for $x_{as}$\;
$Func(x_{as}, e, e_{a})$\;
  Mark that $e$ is examined\;
  }
   Mark $x_{as}$ as examined.
  }
\end{algorithm}

\begin{myeg}
\label{eg:attack-enforced-anonymity-violation-strategy}
(Example~\ref{eg:attack-enforced-anonymity-violation-strategy-func} continued)
Based on the final verifier in Fig.~\ref{fig:final-verifier-2469}, we can obtain an attack-enforced anonymity violation strategy following Algorithm~\ref{alg:systhesis-attack-strategy-algorithm}.

According to Line~1 in Algorithm~\ref{alg:systhesis-attack-strategy-algorithm}, $x_{0,as}=(A,0,\{1,10\})$ and $X_{as}=\{(A,0,\{1,10\})\}$.
By lines~2--4, we choose $e_{a}=N$ as the attack action  for $x_{0,as}$;
 we have $x_{as}=(S,0N,\{1,10\})$ and add $f_{as}(x_{0,as},\varepsilon/N)=x_{as}$ to $f_{as}$, resulting in $X_{as}=\{(A,0,\{1,10\}),(S,0N,\{1,10\})\}$.

We next examine $x_{as}=(S,0N,\{1,10\})$.
For $e=a\in T_{aobs}(x_{as}^{3})\cap E=$ $\{a,d\}$,  we choose $e_{a}=N$ for $x_{as}$.
Then, we have $x_{as}'=(S,0N,\{2,3\})$, $f_{as}(x_{as}$, $a/N)=x_{as}'$, and $X_{as}=\{(A,0,\{1,10\})$, $(S,0N,\{1,10\})$, $(S,0N,\{2,3\})\}$.
Similarly, for $e=d$, we have $x_{as}'=(S,0N,\{6,9\})$, $f_{as}(x_{as},d/N)=x_{as}'$, and $X_{as}=\{(A,0,\{1,10\})$, $(S,0N,\{1,10\})$, $(S,0N,\{2,3\})$, $(S,0N,\{6,9\})\}$.
Now, $(S,0N,\{1,10\})$  is examined.

Next, we examine $x_{as}=(S,0N,\{2,3\})$.
For $e=b,c\in$ $T_{aobs}(x_{as}^{3})\cap E=\{b,c\}$,  we choose $e_{a}=Y$.
Subsequently, we  call $func(x_{as},e,e_{a})=func((S,0N,\{2,3\}),e,Y)$.
According to Algorithm~\ref{alg:call}, 
we have  $x_{fv}=f_{fv}(x_{as}$, $e)=(A,0,\{4,5\})$ and  $x_{fv}'=f_{fv}(x_{fv}$, $e_{a})=(AY,0Y,\{4,5\})$; 
$f_{as}((S,0N,\{2,3\})$, $e/Y0)=(S,1,\{5\})$, $f_{as}((S,0N,\{2,3\})$, $e/Y1)=(S,1,\{4\})$; 
$X_{as}=\{(A,0,\{1,10\})$, $(S,0N,\{1,10\})$, $(S,0N,\{2,3\})$, $(S,0N,\{6,9\})$, $(S,1,\{5\})$, $(S,1,\{4\})\}$.
Finally, $(S,0N,\{2,3\})$  is examined.

Performing the procedures until that all $x_{as}\in X_{as}$ have been examined, the Mealy automaton representing the attack strategy is illustrated in Fig.~\ref{fig:attack-strategy-Mealy}. 
    $\hfill \triangle$
\end{myeg}

\begin{mycom}
Let $x_{fv}\in X_{fv1}$ be the considered state in the final verifier.
For each $e\in T_{aobs}(x_{fv}^{3})$, the intruder should choose the attack action $e_{a}$ for $e$.
If $e_{a}=N$, then there is one new $x_{fv}'\in X_{fv1}$ that should be added to $X_{as}$.
If $e_{a}=Y$, then there are at most two $x_{fv}'\in X_{fv1}$ to be added to $X_{as}$.
There are at most $|E|$ events that should be considered for $x_{fv}\in T(x_{fv})$.
Therefore, the complexity of computing attacked-enforced anonymity violation strategy using Algorithm~\ref{alg:systhesis-attack-strategy-algorithm}  is $O(2|E||X_{fv1}|)$. 
  \end{mycom}

\begin{myeg}
\label{eg:attack-enforced-anonymity-violation-strategy-process}
(Example~\ref{eg:attack-enforced-anonymity-violation-strategy} continued)
An attack-enforced anonymity violation strategy can be obtained in Fig.~\ref{fig:attack-strategy-Mealy}.
The obtained strategy works as follows.
At the initial state $(A,0,\{1,10\})$, if the system generates nothing, the intruder does not perform a state attack. Then,  $(S,0N,\{1,10\})$ is reached, which is denoted by $f_{as}((A,0,\{1,10\}),\varepsilon/N)=(S,0N,\{1,10\})$.
If the system generates $a$ at the initial state, the intruder does not perform a state attack, and then $(S,0N,\{2,3\})$ is reached; this is denoted by $f_{as}((S,0N,\{1,10\}),a/N)=(S,0N,\{2,3\})$.
If the system generates $d$ at the initial state, the intruder does not perform a state attack,  and then $(S,0N,\{6,9\})$ is reached; this is denoted by $f_{as}((S,0N,\{1,10\}),d/N)=(S,0N,\{6,9\})$.

When the system generates $b$ after $a$, the intruder performs a state attack.
If the intruder obtains the attack result ``0", then it knows that $(S,1,\{5\})$ is reached and the current state of the system is 5; this is denoted by $f_{as}((S,0N,\{2,3\}),b/Y0)=(S,1,\{5\})$.
If the intruder obtains the attack result ``1", then it knows that $(S,1,\{4\})$ is reached and the current state of the system is 4; this is denoted by $f_{as}((S,0N,\{2,3\}),b/Y1)=(S,1,\{4\})$.

Following the same procedures, regardless of how the system plays, using the Mealy automaton in Fig.~\ref{fig:attack-strategy-Mealy}, the intruder can decide to perform or not to perform a state attack based on the previous attack actions as well as the corresponding attack results for the aim of  anonymity violation.
    $\hfill \triangle$
\end{myeg}

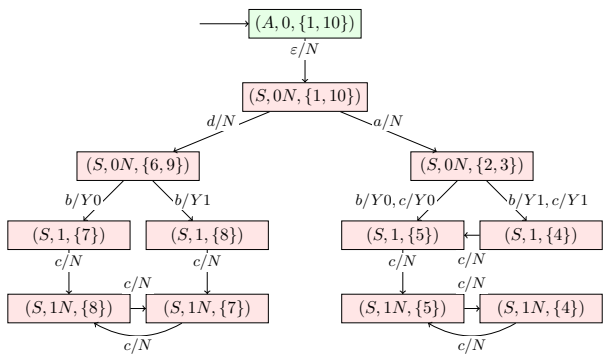
\begin{figure}[!htbp] 
\centering
\scalebox{0.65}{

\begin{tikzpicture}[
 ->,
    node distance=2cm,
  sstate/.style={rectangle, draw, minimum width=2.5cm, minimum height=0.5cm, fill=red!10, align=center},
  astate/.style={rectangle, draw, minimum width=2cm, minimum height=0.5cm, fill=green!10, align=center},
   aystate/.style={rectangle, draw, minimum width=2cm, minimum height=0.5cm, fill=blue!10, align=center},
  emptycell/.style={draw=none, fill=none, minimum width=2cm, minimum height=0.5cm},
  labelstyle/.style={midway, above, font=\small, fill=white, inner sep=1pt},
  initial text=,
  initial distance=1cm
]

  \node[astate,initial] (s1) {$(A,0,\{1,10\})$}; 
  \node[sstate,below of=s1,yshift=0.5cm] (s2) {$(S,0N,\{1,10\})$}; 
  \node[sstate,  below  right of=s2,xshift=2cm] (s3) {$(S,0N,\{2,3\})$}; 
  \node[sstate,  below right of=s3] (s4) {$(S,1,\{4\})$}; 
  
\node[sstate, below left of=s2,xshift=-2cm] (s5) {$(S,0N,\{6,9\})$}; 
  \node[sstate, below left of=s5] (s6) {$(S,1,\{7\})$}; 
  \node[sstate,  below  right of=s5] (s7) {$(S,1,\{8\})$}; 
  \node[sstate,below left of=s3] (s8) {$(S,1,\{5\})$}; 
  
  \node[sstate,below of=s6,yshift=0.5cm] (s9) {$(S,1N,\{8\})$}; 
  \node[sstate,below of=s7,yshift=0.5cm] (s10) {$(S,1N,\{7\})$}; 
  \node[sstate,below of=s8,yshift=0.5cm] (s11) {$(S,1N,\{5\})$}; 
  \node[sstate,below  of=s4,yshift=0.5cm] (s12) {$(S,1N,\{4\})$};

\draw[->] (s1) -- node[labelstyle] {$\varepsilon/N$} (s2);

\draw[->] (s4) -- node[labelstyle,below,yshift=-0.3cm] {$c/N$} (s8);
\draw[->] (s8) -- node[labelstyle] {$c/N$} (s11);
\draw[->] (s11) -- node[labelstyle,above,yshift=0.3cm] {$c/N$} (s12);
\draw[->,bend left] (s12) to node[labelstyle,below] {$c/N$} (s11);
\draw[->] (s6) -- node[labelstyle] {$c/N$} (s9);
\draw[->] (s7) -- node[labelstyle] {$c/N$} (s10);
\draw[->] (s9) -- node[labelstyle,above,yshift=0.3cm] {$c/N$} (s10);
\draw[->,bend left] (s10) to node[labelstyle,below] {$c/N$} (s9);
\draw[->] (s2) -- node[labelstyle] {$a/N$} (s3);
\draw[->] (s2) -- node[labelstyle] {$d/N$} (s5);
\draw[->] (s3) -- node[labelstyle,right] {$b/Y1,c/Y1$} (s4);
\draw[->] (s3) -- node[labelstyle,left] {$b/Y0,c/Y0$} (s8);
\draw[->] (s5) to node[labelstyle,right] {$b/Y1$} (s7);
\draw[->] (s5) to node[labelstyle,left] {$b/Y0$} (s6);

\end{tikzpicture}
}
\caption{\label{fig:attack-strategy-Mealy} Attack-enforced anonymity violation strategy  for the NFA in Fig.~\ref{fig:system} under $X_{A}=\{2,4,8,9\}$ and $D=1$.}
\end{figure}

\begin{myrek}
  When focusing on opacity violations,  after predefining the set of secret states $X_{S}$, $|x_{aobs}^{3}|= 1$ is replaced with $x_{aobs}^{3}\subseteq X_{S}$ in the corresponding definitions and algorithms.
\end{myrek}

\section{Conclusions}

In this article, we consider an intruder that has full knowledge of the system and can perform a limited number of state attacks to obtain the membership of the possible current state to the set of attacked states.
Focusing on fully-observable nondeterministic systems, we develop algorithms and constructions to verify whether the intruder can become aware of the exact state of the system based on the sequence of observations it received and the corresponding sequence of performed state attacks. 

In order to capture the ability of the intruder to use a bounded number of attacks to force the system to reach an anonymity-violating state regardless of how  the system behaves, the notion of attack-enforced anonymity violation is proposed.
We also develop an algorithm to verify whether the system satisfies attack-enforced anonymity violation if the system is determined to be anonymity-violating.
Furthermore, an algorithm for the intruder to synthesize an attack-enforced anonymity violation strategy is presented, illustrating how the state attack operations should be performed as the system evolves. 

Future work will consider scenarios where an intruder is capable of launching multiple state attacks, meaning the intruder can access various sets of attacked states and dynamically select combinations of such attacks.
Another interesting line   would be to investigate the synthesis of attack strategies for anonymity-violation under scenarios where the intruder can  perform both state attacks and actuator attacks (i.e., disable events that are enabled or enable events that are disabled).

\begin{IEEEbiography}[{\includegraphics[width=1in,height=1.15in,clip,keepaspectratio]{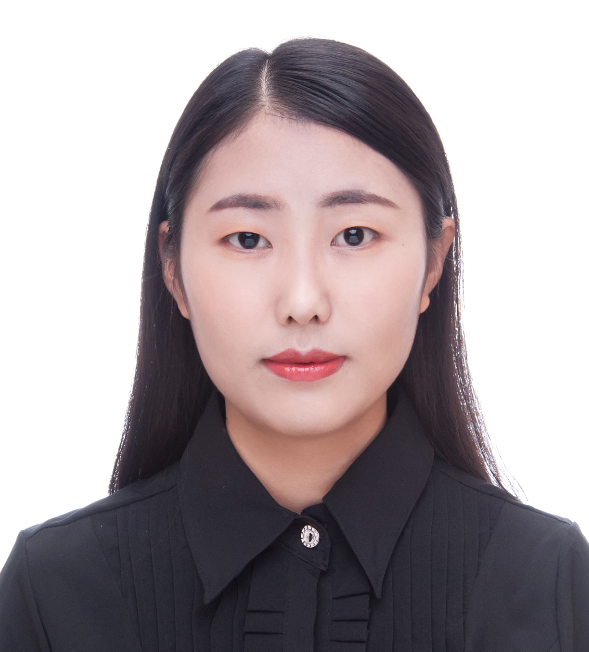}}]{Xiaoyan Li} received the B.S. degree in automation
from Taiyuan University of Technology,
Taiyuan, China, in 2017, and  the  Ph.D.  degree in control theory and control
engineering from Xidian University, Xi'an, China, in 2022.
She is currently with the School of Information and Communication Engineering, North University of China.
Her research interests include Petri net theory, robust supervisory control of
automated manufacturing systems, and security and privacy of cyber-physical systems.
\end{IEEEbiography}

\begin{IEEEbiography}[{\includegraphics[width=1in,height=1.15in,clip,keepaspectratio]{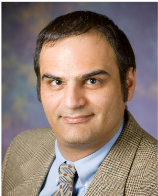}}]{Christoforos N. Hadjicostis} (Fellow, IEEE) received the B.S.  degrees  in  electrical  engineering, in computer science and engineering, and in mathematics in 1993, the M.Eng. degree in electrical engineering and computer  science  in  1995,  and  the  Ph.D.  degree  in electrical engineering and computer science in 1999, all  from  the  Massachusetts  Institute  of  Technology, Cambridge. 
Since  2007,  he  is a  Professor with  the  Department  of  Electrical  and Computer Engineering, University of Cyprus. 
His research focuses on fault diagnosis and tolerance in distributed dynamic systems,  error  control  coding,  monitoring,  diagnosis  and  control  of  large-scale  discrete-event  systems,  and  applications  to  network  security,  anomaly detection, energy distribution systems, and medical diagnosis.
Dr.  Hadjicostis  serves  as  Editor in Chief  of the  Journal  of  Discrete  Event  Dynamic  Systems  and  as Senior  Editor  of  IEEE  Transactions  on  Automatic Control.
\end{IEEEbiography}

\end{document}